\documentclass[a4paper,onecolumn,11pt,accepted=2024-09-25]{quantumarticle}
\pdfoutput=1

\usepackage[utf8]{inputenc}
\usepackage[english]{babel}
\usepackage{hyperref}
\usepackage{tikz}
\usepackage{lipsum}

\usepackage{amsmath}
\usepackage{amsfonts}
\usepackage{amsthm}
\usepackage{amssymb}
\usepackage{graphicx}
\usepackage{enumerate}
\usepackage{color}
\usepackage[T1]{fontenc}
\usepackage{braket}
\usepackage{todonotes}
\usepackage{soul}
\usepackage{subfigure}
\usepackage{mathrsfs}
\usepackage{float}
\usepackage{bm}
\usepackage{multirow}

\usepackage{pifont}
\usepackage[makeroom]{cancel}

\graphicspath{{figures/}}

\DeclareMathOperator{\Tr}{Tr}

\newtheorem{theorem}{Theorem}
\newtheorem{lemma}{Lemma}
\newtheorem{corollary}{Corollary}
\newtheorem{proposition}{Proposition}

\newtheorem*{theorem*}{Theorem}
\newtheorem{definition}{Definition}
\newtheorem*{proposition*}{Proposition}
\newtheorem{examp}{Example}

\newcommand{\G}{\mathcal G}
\def\>{\rangle}
\def\<{\langle}

\begin{document}

\title{(Almost-)Quantum Bell Inequalities and Device-Independent Applications}
\author{Yuan Liu}
\email{yuan59@connect.hku.hk}
\affiliation{School of Computing and Data Science, The University of Hong Kong, Pokfulam Road, Hong Kong}

\author{Ho Yiu Chung}
\email{hollis\_chung@yahoo.com}
\affiliation{School of Computing and Data Science, The University of Hong Kong, Pokfulam Road, Hong Kong}

\author{Ravishankar Ramanathan}
\email{ravi@cs.hku.hk}
\affiliation{School of Computing and Data Science, The University of Hong Kong, Pokfulam Road, Hong Kong}
\orcid{0000-0003-1119-8721}

\begin{abstract}
 Investigations of the boundary of the quantum correlation set have gained increased attention in recent years. This is done through the derivation of quantum Bell inequalities, which are related to Tsirelson's problem and have significant applications in device-independent (DI) information processing. 
However, determining quantum Bell inequalities is a notoriously difficult task and only isolated examples are known. In this paper, we present families of (almost-)quantum Bell inequalities and highlight four foundational and DI applications. 
Firstly, it is known that quantum correlations on the non-signaling boundary are of crucial importance in the task of DI randomness extraction from weak sources. In the practical Bell scenario of two players with two $k$-outcome measurements, we derive quantum Bell inequalities that demonstrate a separation between the quantum boundary and certain portions of the boundaries of the no-signaling polytope of dimension up to $4k-8$, extending previous results from nonlocality distillation and the collapse of communication complexity. 
Secondly as an immediate by-product, we give a general proof of Aumann’s Agreement theorem for quantum systems as well as the almost-quantum correlations, which implies Aumann’s agreement theorem is a reasonable physical principle in the context of epistemics to pick out both quantum theory and almost-quantum correlations from general no-signaling theories.
Thirdly, we present a family of quantum Bell inequalities in the two players with $m$ binary measurements scenarios, that we prove serve to self-test the two-qubit singlet and the corresponding $2m$ measurements. Interestingly, this claim generalizes the result for $m=2$ discovered by Tsirelson-Landau-Masanes and shows an improvement over the state-of-the-art Device-Independent Randomness-Amplification (DIRA).
Lastly, we use our quantum Bell inequalities to derive the general form of the principle of no advantage in nonlocal computation, which is an information-theoretic principle that serves to characterize the quantum correlation set. 
\end{abstract}
\maketitle
\newpage

\section{Introduction}
	One of the most striking features of quantum mechanics is nonlocality, the phenomenon of violation of Bell inequalities by separated physical systems. The correlations between local measurement outcomes on such systems show, in a fully device-independent manner, that quantum theory differs fundamentally from all classical theories that are constrained by the principle of local causality \cite{bell1975,bell1995free}. Besides their foundational interest, in recent years, the quantum correlations have been shown to be a vital resource in device-independent (DI) information processing applications, such as quantum key distribution \cite{barrett2005no,acin2007device}, randomness extraction and expansion \cite{pironio2010random,pironio2013security}, self-testing of quantum states and measurements \cite{mayers1998quantum,mayers2003self}, and reduction of communication complexity \cite{buhrman2010nonlocality}. 
	
	The Bell inequalities delineate the boundary of the classical correlation set, and any violation of a Bell inequality indicates that the observed distribution is nonlocal. Moreover, the verification of nonlocal correlations (and the correct execution of DI tasks built upon these) can be performed by simple statistical tests of the measurement devices and a fundamental rule of nature, viz. the no-superluminal signaling principle of relativity. While the classification of the entire set of Bell inequalities for arbitrary numbers of measurement systems, inputs and outputs is a challenge, at least a systematic method for the identification of novel Bell inequalities is known since the work by Pitowsky \cite{pitowsky1991correlation}. 

    On the other hand, the quantum correlation set, denoted $\mathrm{Q}$, is known to lie in between the classical set $\mathrm{L}$ and the general no-signaling set $\mathrm{NS}$~\cite{popescu1994quantum}. 
    It's essential to more precisely characterize the quantum correlations set, and understand where it stands between the classical and general non-signaling correlations polytopes. For instance, this characterization is crucial to identify the optimal quantum correlations for different applications. 
    However, the task of characterizing the quantum correlation set has proven to be a much more challenging task compared to the classical one ~\cite{goh2018geometry}. Indeed, for a given correlation there is no general method to determine whether it belongs to the quantum correlation set, no need to mention the difficulty of finding the quantum system that saturates this correlation. Several different approaches have been proposed in the literature from different perspectives in an attempt to partially answer this question. For instance, the NPA hierarchy~\cite{navascues2008convergent,navascues2007bounding} provides a numerical approximation of the quantum correlation set from the outside, the convexity of the quantum correlation set and its extremal points (for some specific scenarios) have been studied in~\cite{masanes2003necessary,goh2018geometry,ishizaka2018necessary,mikos2023extremal}. It has been proven that for the scenarios involving any number of parties with binary inputs, the classical correlation polytope is dual to the no-signaling polytope~\cite{fritz2012polyhedral}, the duality of the quantum correlation set, which is convex but is in general not a polytope unlike $\mathrm{L}$ and $\mathrm{NS}$, is less known and has been studied for some specific scenarios~\cite{cirel1980quantum,landau1988empirical,masanes2003necessary,fritz2012polyhedral,le2023quantum,victor2024extremal}. 
    Analogs to using Bell inequalities to delineate the boundary of the classical correlation set, one can also derive the quantum Bell inequality to study the boundary of the quantum correlation set, however, these quantum Bell inequalities are more complicated compared to Bell inequalities, usually non-linear and only a few examples have been found so far \cite{rai2019geometry,chen2022quantum,almeida2010guess,ramanathan2014characterizing,ramanathan2021violation,linden2007quantum,le2023quantum,sun2018uncertainty}.

    Even though much effort has been put in, the boundary of the quantum correlation set is still not completely understood. One question that is highly related to the characterization of the quantum correlation set is to exclude the no-signaling correlations (usually termed as no-signaling boxes) from the quantum set. It's proven by one of us in \cite{ramanathan2016no} that all nonlocal vertices of the no-signaling polytope are not quantum realizable, this result is pretty general since it holds for any Bell scenario with the arbitrary number of parties, inputs and outputs. Apart from that, in the simplest Bell scenario, the authors in~\cite{chen2022quantum} looked for all the boxes on the coincident boundary of the quantum correlation set and no-signaling set, 
    and the nonlocal boxes that collapse communication complexity~\cite{proulx2023extending} are also outside of the quantum set. 
    It has been observed that the physically achievable boxes set must be closed under wirings~\cite{allcock2009closed}. In~\cite{beigi2015monotone}, the authors provided tools to systematically construct sets of non-local boxes that are closed under wirings and by applying these tools to the noisy PR box (isotropic boxes), they found a continuum of sets of non-local boxes that are closed under wirings, thus partially answering the question of finding the feasible non-local boxes in nature. Afterward, also in the simplest Bell scenario, the authors~\cite{rai2019geometry,brito2019nonlocality} identified a board class of  `quantum void' in the no-signaling set by utilizing the nonlocality distillation protocol.

	The practical implications of Quantum Bell Inequalities are profound, particularly in the realm of DI applications. One important task that has gained prominence in recent years is self-testing~\cite{mckague2012robust}, namely the unique identification (up to local isometries) of a quantum state and measurements, solely from the observed correlations in a Bell test without trusting the internal workings of the devices. As such, this task requires the identification of quantum correlations that can be generated in such a unique manner. The study of Quantum Bell Inequalities is also important from a fundamental viewpoint in the problem of identifying appropriate physical and information-theoretic principles that single out the set of quantum correlations from amongst general no-signaling ones. Of particular importance are the principle of information causality \cite{pawlowski2009information}, macroscopic locality~\cite{navascues2010glance}, local orthogonality \cite{fritz2013local}, no advantage in nonlocal computation \cite{linden2007quantum}, the collapse of communication complexity \cite{brassard2006limit}, as well as Aumann's Agreement theorem which is recently generalized to the quantum system~\cite{contreras2021observers}, all of which have been shown to lead to non-trivial bounds on the set of quantum correlations. The identification of nonlocal no-signaling boxes that are excluded from the quantum set serves as a useful testing ground and pointers towards the ultimate principle picking out the quantum set.

In this paper, we explore the boundary of the quantum correlation set with specific regard to regions coinciding with a no-signaling or a local boundary, and non-trivial regions leading to self-testing. To do this, we expand on a class of non-linear (Almost-)Quantum Bell Inequalities defining the boundary of the almost-quantum Set \cite{navascues2015almost}. Such inequalities were used to exclude all nonlocal vertices of the no-signaling polytope (for arbitrary number of parties, inputs and outputs) by one of us in \cite{ramanathan2016no}. Here, we explore these inequalities to exclude further non-trivial regions of the no-signaling polytope. Specifically, in the $(2,2,k)$ Bell scenario (with two players performing two $k-$nary measurements), we derive optimal inequalities that show the exclusion of nonlocal correlations on certain portions of the boundaries of the no-signaling polytope of dimension up to $4k-8$. This extends the known region of excluded boxes from the no-signaling boundary obtained in other approaches introduced before.
As a direct consequence, we prove that Aumann's Agreement theorem holds for the quantum systems as well as the almost-quantum correlations. This result not only corrects the proof presented in~\cite{contreras2021observers} but also answers the open question that Aumann's Agreement theorem is a reasonable physical principle in the context of epistemics and a common feature of both the quantum thoery and almost-quantum correlations.
Secondly, we derive a class of tight quantum Bell inequalities in the $(2,m,2)$ Bell scenario (with two players performing $m$ binary measurements) and show their usefulness in self-testing the two-qubit singlet state. In this regard, we generalize the results regarding the self-testings of the singlet in the $(2,2,2)$ scenario obtained in \cite{wang2016all} and the self-testing of the correlations leading to the optimal violation of the chained Bell inequality in \cite{vsupic2016self}. 
Finally, we study the faces of the $(2,m,2)$ correlation set (excluding the local marginals), and identify low-dimensional regions in which the quantum correlation set coincides with the classical correlation polytope. In this regard, we generalize the results obtained in \cite{linden2007quantum}.

\section{Almost-Quantum Bell Inequalities}
\subsubsection{Bell scenario and the set of quantum correlations}
We label a Bell scenario as $(n,m,k)$, in which $n$ space-like separated parties perform measurements on a shared physical system, each party has $m$ choices of local measurement and each measurement yields $k$ possible outcomes. Specifically, we focus on the bipartite scenario $(2,m,k)$ (experimenters Alice and Bob), in which we label the local measurements as $x,y\in[m]$ (where $[m] =\{1,2,\ldots,m\}$) and the possible outcomes of each measurement are $a,b\in[k]$ (where $[k]=\{1,2\ldots,k\}$) respectively. 
Let $p(a,b|x,y)$ denote the probability of obtaining the outcomes $a,b$ given that Alice and Bob chose measurement settings $x,y$, these probabilities obey non-negativity $p(a,b|x,y)\geq 0,\forall x,y,a,b$ and are normalized $\sum _{a,b}p(a,b|x,y)=1,\forall x,y$.
A box then refers to a collection of conditional probability distributions $\mathrm{P}:=\{p(a,b|x,y)\}_{x,y\in[m]; a,b\in[k]}$ (we will also write a box in terms of a vector $\vec{\mathrm{P}}$ or $|\mathrm{P}\>$). There are three different physical models of our interest which can be translated into three different types of constraints on the boxes.
The first set of interest consists of general no-signaling boxes. This set is defined by the requirement that the local marginal probabilities of one party are well defined and not affected by the input of the other party, i.e., the experimenters cannot signal their inputs to each other.  
\begin{equation}
	\begin{split}
		\sum_a p(a,b|x,y)= \sum_a p(a,b|x',y) \qquad \forall b,x,x',y\\
		\sum_b p(a,b|x,y)= \sum_b p(a,b|x,y') \qquad \forall a,x,y,y'
	\end{split}
\end{equation}
The set of all boxes satisfying the above no-signaling constraints forms the no-signaling convex polytope $\mathrm{NS}(2,m,k)$.
The second set of interest consists of the Quantum boxes. In this case, there exists a quantum state $\rho$ in some Hilbert space and a set of measurement operators (positive operator valued measure (POVM) elements) $\{A_x^a\}$ and $\{B_y^b\}$ such that the joint conditional probabilities obey the Born's rule:
\begin{equation}
	p(a,b|x,y)= \Tr(\rho A_x^a\otimes B_y^b) \qquad \forall a,b,x,y
\end{equation}
All such quantum boxes form the quantum convex set $\mathrm{Q}(2,m,k)$.
The third set of interest consists of the Local Causal (Classical) boxes. In this case, there exist hidden variables $\lambda$ with distribution $p(\lambda)$, and corresponding local response functions $p(a|x,\lambda),P(b|y,\lambda)$ such that 
\begin{equation}
	p(a,b|x,y)= \sum_{\lambda} P(\lambda)p(a|x,\lambda)p(b|y,\lambda) \qquad \forall a,b,x,y
\end{equation}
We denote by $\mathrm{L}(2,m,k)$ the local convex polytope, and the extreme points of this set $\mathrm{L}(2,m,k)$ are the local deterministic boxes. In general, the relationship of these three sets is $\mathrm{L}(n,m,k)\subsetneq \mathrm{Q}(n,m,k)\subsetneq \mathrm{NS}(n,m,k)$.

\subsubsection{Almost-quantum set and the Theta body of the Orthogonality Graph}

In ~\cite{navascues2008convergent,navascues2007bounding}, Navascues, Pironio, and Acín introduced a hierarchy of semi-definite programs (SDPs) that form an outer approximation of the quantum set. In the SDPs, the requirement of tensor product structure of measurement operators of space-like separated parties is replaced with the commutation requirement of the operators, thus the conditional probabilities are expressed as
\begin{equation}
	p(a,b|x,y)= \<\psi |A_x^{a} B_y^{b}|\psi\> \qquad \forall a,b,x,y
\end{equation}
with projection operators $\{A_x^{a}\},\{B_y^{b}\}$ and the commutation requirement $[A_x^{a},B_y^{b}]=0,\forall a,b,$ $x,y$. In the hierarchy, let $\mathrm{S}_1:=\{\mathbb{I}\}\cup\{A_x^{a}\}\cup\{B_y^{b}\}$, and the set $\mathrm{S}_k,k> 1$  contains $S_{k-1}$ and products of $k$ elements of $\mathrm{S}_1$. For example, $\mathrm{S}_1=\{\mathbb{I}\}\cup\{A_x^{a}\}\cup\{B_y^{b}\}$, $\mathrm{S}_2=\mathrm{S}_1 \cup \{A_x^{a}A_{x'}^{a'}\} \cup \{B_y^{b}B_{y'}^{b'}\} \cup \{A_x^{a}B_{y}^{b}\}$, etc. The moment matrix $\Gamma^{(k)}$ is associated with the set $\mathrm{S}_k$, the entry of moment matrix is $\Gamma^{(k)}_{i,j}=\<\psi| s_i^{(k)\dag}s_j^{(k)}|\psi\>$ where $s_i^{(k)}\in \mathrm{S}_k, \forall i\in[|\mathrm{S}_k|]$ and the size of the moment matrix is $|\mathrm{S}_k|\times |\mathrm{S}_k|$. It's clear that the moment matrix $\Gamma^{(k)}$ is positive semi-definite $\Gamma^{(k)}\succeq 0,\forall k.$
A box $\mathrm{P}$ belongs to the set $\mathrm{Q}_k$, if the elements of the box $\mathrm{P}$ are equal to the corresponding entries of the moment matrix $\Gamma^{(k)}$. The sets $\mathrm{Q}_k$ are convex and they converge toward the quantum set $\mathrm{Q}$ from outside $\mathrm{Q}_1\supsetneq\mathrm{Q}_2\supsetneq\mathrm{Q}_3\supsetneq\cdots \supsetneq\mathrm{Q}$. Thus, for a box $\mathrm{P}$, the lack of existence of $\Gamma^{(k)}\succeq 0$ at any level $k$ of the hierarchy means $\mathrm{P}$ is outside of the quantum set. A particular level of the hierarchy between levels $1$ and $2$ is of great interest, which is labeled as level $1+ab$. And this set has been highlighted as the \textit{Almost-Quantum} set $\widetilde{\mathrm{Q}}$~~\cite{navascues2015almost}, in the sense that the set $\widetilde{\mathrm{Q}}$ satisfies almost all the reasonable information-theoretic principles (except the principle of information causality, for which a proof is not known), that pick out quantum correlations from general no-signaling ones.

The orthogonality graph $G_B$~\cite{sainz2014exploring,cabello2010non} can be constructed for any Bell scenario. Take the Bell scenario $(2,m,k)$ as an example, each event $(a,b|x,y)$ in the Bell scenario $(2,m,k)$ corresponds to a distinct vertex of the graph, and two vertices are connected by an edge if the two events involve different outcomes for the same local measurement by at least one of the parties (such events are termed \textit{locally orthogonal}). The Theta body $\text{TH}(G)$ of any graph $G=(V,E)$ is the convex set defined as follows.
\begin{definition}(\cite{grotschel1986relaxations}) For any graph $G=(V,E)$, $\text{TH}(G):=\{\vec{\mathrm{P}}=(|\<\psi|u_i\>|^2:i\in |V|)\in\mathbb{R}_+^{|V|}:\||\psi\>\|=\||u_i\>\|=1,\{|u_i\>\} \text{ is an orthonormal representation of }G\}$.   
\end{definition}
We consider the orthogonality graph $G_B$ for a Bell scenario, and define a subset of the Theta body $\text{TH}^{(c)}(G_B)\subsetneq \text{TH}(G_B)$, in which additional clique constraints are applied to the maximum cliques $\{c\}$ of the orthogonality graph $G_B$ as:
\begin{definition}(\cite{cabello2010non})
	For the orthogonality graph $G_B$ corresponding to the Bell scenario $(n, m, k)$, define the set $\text{TH}^{(c)}(G)$ as 
 \begin{equation}
     \text{TH}^{(c)}(G_B):=\left\{\vec{\mathrm{P}}=\left(\left|\left\langle\psi | u_i\right\rangle\right|^2\right) \in \text{TH}(G_B):  \sum_{i \in c}\left|\left\langle\psi |u_i\right\rangle\right|^2=1,\forall c \in C_{n, n s}\right\}.
 \end{equation}
 Here, $C_{n, n s}$ denotes the set of maximum cliques of orthogonality graph $G_B$.
\end{definition}
It has been shown that the set $\text{TH}^{(c)}(G_B)$ is equivalent to the almost-quantum set. 
\begin{theorem*}(\cite{acin2015combinatorial})
	For any Bell scenario $(n,m,k)$, the almost-quantum set is equivalent to the Theta body with maximum clique constraints, i.e., $\widetilde{\mathrm{Q}}= \text{TH}^{(c)}(G_B)$.
\end{theorem*}
In summary, for a Bell scenario $(2,m,k)$, we get the following important relationships:
\begin{equation}
	\text{TH}(G_B)\supsetneq \text{TH}^{(c)}(G_B)=\widetilde{\mathrm{Q}}=\mathrm{Q}_{1+ab}\supsetneq \mathrm{Q}
\end{equation}
Thus, the exclusion of a box $\mathrm{P}$ from the set $\text{TH}(G_B)$ implies that this box has no quantum realization, i.e., $\mathrm{P}\notin \mathrm{Q}(2,m,k) $. Now, we utilize the following dual characterization of the set $\text{TH}(G_B)$~\cite{fujie2002}: 
\begin{equation}
	\label{eq:quantum-Bell}
	\text{TH}(G_B)=\{|P\>\in\mathbb{R}^{|V|}:\<P|M|P\>-\sum_{i\in|v|}M_{i,i}|P\>_i\leq 0, \forall M\in\mathbb{M}\}
\end{equation}
with 
\begin{equation}
	\label{eq:M-bound}
	\mathbb{M}:=\{M\in \mathbb{S}^{|V|}: M_{u,v}=0 (u\neq v, u\nsim v) , M\succeq 0\}.
\end{equation}
The set of inequalities $\<P|M|P\>-\sum_{i\in|v|}M_{i,i}|P\>_i\leq 0$ therefore defines a set of almost-quantum Bell inequalities for any number of players, inputs and outputs separating (almost-) quantum from the post-quantum set. Choosing appropriate $M$ satisfying \eqref{eq:M-bound} allows us to identify non-trivial boundaries of the almost-quantum set (and recovers some of the boundaries identified by principles such as macroscopic locality \cite{gachechiladze2022quantum,yang2011quantum} or local orthogonality \cite{fritz2013local}). In certain Bell scenarios such as when two players perform binary measurements, this class of inequalities also allows to identify tight boundary regions of the quantum set.

\section{Excluding nonlocal no-signaling Boxes from the Quantum Set}\label{sec_exclude}
We have seen a class of inequalities that provide a boundary on the (almost-)quantum set. In ~\cite{ramanathan2016no}, the authors used these inequalities to exclude, from the quantum set, all nonlocal vertices of the no-signaling polytope for any number of players, inputs and outputs. In this section, we generalize this statement and identify optimal quantum Bell inequalities to exclude specific nonlocal regions of the no-signaling polytope. We thereby extend known results regarding the excluded region of the no-signaling polytope, that were previously obtained using the method of nonlocality distillation (any box that can be distilled to a PR box through local operations and shared randomness must be excluded from the quantum set) \cite{rai2019geometry,brito2019nonlocality} and the collapse of communication complexity \cite{proulx2023extending}. The purpose of this section is more generally to illustrate the utility of the class of inequalities in \eqref{eq:quantum-Bell} in easily exclude nonlocal regions of the no-signaling polytope from the quantum set.

To identify the optimal $M$ satisfying \eqref{eq:M-bound} that serves to exclude a nonlocal box $\mathrm{P}$, we solve the following SDP:
\begin{align}
\begin{split}
	\max_M\quad & \<\mathrm{P}|M|\mathrm{P}\>-\sum_{i\in|V|}M_{i,i}|\mathrm{P}\>_i\\
	s.t. \quad & M\succeq 0\\
	& M_{u,v}=0 (u\neq v, u\nsim v)\\
	&M\in \mathbb{S}^{|V|}
\end{split}
\end{align}
where $|\mathrm{P}\>$ is the vector form of the box $\mathrm{P}$. If the solution to the SDP is a strictly positive value, then the corresponding quantum Bell inequality certifies that the box $\mathrm{P}$ is not in the quantum set.
In this section, for specific regions of the no-signaling boundary in the $(2,2,2)$ and the $(2,2,k)$ Bell scenarios we analytically present psd matrices $\widetilde{M}$ satisfying  \eqref{eq:M-bound} such that $\<\mathrm{P}|\widetilde{M}|\mathrm{P}\>-\sum_{i\in|v|}\widetilde{M}_{i,i}|\mathrm{P}\>_i>0$, thereby excluding these regions from the quantum set.

\begin{theorem} 
\label{theo1}
In the $(2,2,2)$ Bell scenario, let $\mathrm{P}$ be a point on a face of the no-signaling polytope $\mathrm{NS}(2,2,2)$ of dimension $d\leq 4$, such that $\mathrm{P}\notin \mathrm{L}(2,2,2)$, then $\mathrm{P}\notin \mathrm{Q}(2,2,2)$.
\end{theorem}

It's well-known that in this simplest $(2,2,2)$ Bell scenario, there are $24$ extreme vertices of $\mathrm{NS}(2,2,2)$ and $8$ of them are nonlocal denoted as PR boxes and the other $16$ are local deterministic boxes. The PR boxes~\cite{popescu1994quantum} are equivalent up to relabeling of the inputs and outputs. Due to the symmetries of the $(2,2,2)$ scenario, each PR box is neighboring to $8$ local deterministic boxes and they form an $8$-dimensional simplex~\cite{barrett2005nonlocal}, to understand the behavior of boxes on the boundary of the no-signaling polytope, we focus on the boxes on the faces of this simplex. 
For the sake of readability, we provide the proofs for $d=0, 1$ here, while the detailed proofs of other faces are deferred to the Appendix ~\ref{Asec1}. The main idea is that we present quantum Bell inequalities associated with $\widetilde{M}$ for any nonlocal box on the faces of the simplex up to dimension $4$ which indicates that all nonlocal boxes on the corresponding faces are outside of the quantum correlation set.

\begin{proof}

In the proof, we relabel the entries of $|\mathrm{P}\rangle$ which is the vector form of the box $\{p(a,b|x,y)\}$, such that the first five entries of $|\mathrm{P}\rangle$ correspond to the probability of events $(1,1|0,1),(0,0|0,0),$ $(1,1|0,0),(0,0|1,0),(1,0|1,1)$.
\begin{enumerate}
	\item[I.] $0$-dimensional faces (Nonlocal Vertices).\\
	Since the 8 PR boxes in $(2,2,2)$ scenario are equivalent (up to relabeling the indexes of inputs and outputs), we will just analyze one with $p(a,b|x,y) = 1/2$ for $a \oplus b = x \cdot y$. We construct the matrix $M_0$ as:
	\begin{equation}
        \begin{split}
		M_0&=\sum_{i=1}^5 |\iota_{(i,i+1)}\rangle\langle \iota_{(i,i+1)}| =|\iota_{(1,2)}\rangle\langle \iota_{(1,2)}|+| \iota_{(2,3)}\rangle\langle \iota_{(2,3)}| \\
        &+|\iota_{(3,4)}\rangle\langle \iota_{(3,4)}|+ |\iota_{(4,5)}\rangle\langle \iota_{(4,5)}|+ |\iota_{(5,1)}\rangle\langle \iota_{(5,1)}|\succ 0
        \end{split}
	\end{equation}
	where $|\iota_{(i,i+1)}\rangle$ is an indicator (with $0/1$-entries) vector of length 5, with non-zero entries in the positions $i$ and ${i+1}$. $M_0$ is positive definite, so there exists an $\epsilon>0$, such that $\widetilde{M}_0=M_0-\epsilon |\iota_{(1)}\rangle \langle \iota_{(1)}|\succeq 0$. We therefore obtain the inequality 
	\begin{equation}\label{exclude}
		\langle \mathrm{P}|\widetilde{M}_0 |\mathrm{P}\rangle-\sum_{i=1}^5 \left(\widetilde{M}_{0}\right)_{i,i}|\mathrm{P}\>_i=-\epsilon |\mathrm{P}\rangle_1(|\mathrm{P}\rangle_1-1)>0,
	\end{equation}
    recovering the well-known fact that the PR box $\mathrm{P}$ has no quantum realization. 
	Note that here $|\mathrm{P}\>$ is a vector of length $16$ and the $M$ matrix in the definition of Theta body $\text{TH}(G)$ in this scenario is of size $16\times 16$, but we have only picked five events and constructed the matrix $\widetilde{M}_0$ of size $5\times 5$. So that we implicitly extend the matrix $\widetilde{M}_0$ with suitable $0$ entries to obtain \eqref{exclude}. 
	
\item [II.] 1-dimensional faces.
These are the convex combination of a PR box and one neighboring Local deterministic box, that is
	\begin{equation}\label{face1}
		\mathrm{P}=c_{NS}\cdot PR+(1-c_{NS})\cdot L_i
	\end{equation}
	where $i\in \{1,2\ldots,8\}$ and $0<c_{NS}<1$. Without loss of the generality, here we assume the local box in Eq.~\eqref{face1} is the one that $p(0|x)=1,p(0|y)=1,\forall x,y\in\{0,1\}$.
    In this case, we construct a matrix $\widetilde{M}_1$ as:
	\begin{equation}\label{1bchain}
		\widetilde{M}_1=4\cdot \sum_{i=1}^4 | \iota_{(i,i+1)} \rangle \langle \iota_{(i,i+1)}| + | \iota_{(1)} \rangle \langle \iota_{(5)}| + |\iota_{(5)} \rangle \langle \iota_{(1)}| =\left(
		\begin{array}{ccccc}
			4 & 4 & 0 & 0 & 1 \\ 
			4 & 8 & 4 & 0 & 0 \\ 
			0 & 4 & 8 & 4 & 0 \\ 
			0 & 0 & 4 & 8 & 4 \\ 
			1 & 0 & 0 & 4 & 4
		\end{array}\right)
	\end{equation}
	which can be verified to be positive definite by applying elementary row operations to transform $\widetilde{M}_1$ to an upper triangular matrix with all the diagonal entries positive. 
    We obtain the inequality
	\begin{equation}
		\langle \mathrm{P}|\widetilde{M}_1 |\mathrm{P}\rangle-\sum_{i=1}^5 \left(\widetilde{M}_{1}\right)_{(i,i)}|\mathrm{P}\>_i=2|\mathrm{P}\>_1\cdot |\mathrm{P}\>_5>0.
	\end{equation} 
    Thus all the boxes on the 1-dimensional faces of the no-signaling polytope $\mathrm{NS}(2,2,2)$ are excluded from the quantum set $\mathrm{Q}(2,2,2)$ in the same manner.
 \end{enumerate}
 \end{proof}

Note that the optimal quantum point for the well-known Hardy paradox shows that the quantum set reaches a five-dimensional nonlocal boundary of the non-signaling set, so that the theorem is optimal. We now proceed to extend Thm.~\ref{theo1} to more general Bell scenarios $(2,2,k)$ for $k\geq 2$.
To do this, we first study the structure of the no-signaling polytope $\mathrm{NS}(2,2,k)$, specifically focusing on identifying the regions that connect a non-local vertex with certain local vertices.

\begin{lemma}(\cite{barrett2005nonlocal})\label{lem_prk}
The nonlocal vertices of $\mathrm{NS}(2,2,k)$ for two inputs $x,y\in\{0,1\}$ and outputs $a\in\{0,\ldots, k-1\}$ and $b\in\{0,\ldots,k-1\}$ are equivalent under local relabelling to
\begin{equation}\label{eq_prk}
	p{(a, b | x, y)}=\left\{\begin{aligned}
		1 / d:&\quad (b-a) \bmod d=x\cdot y \\
		& \quad a, b \in\{0, \ldots, d-1\} \\
		0: &\quad  \text { otherwise }
	\end{aligned}\right.
\end{equation}
for each $d \in\left\{2, \ldots, k\right\}$.
\end{lemma}
Similar to the case of the PR box in $(2,2,2)$ Bell scenario, all nonlocal vertices in the above lemma with $d=k$ are equivalent up to relabeling, and without loss of generality we will only focus on the one listed above in Eq.~\eqref{eq_prk} with $d=k$, and denote it by $PR^{(k)}$.

We now define a nonlocal game $g_k$ associated with the $PR^{(k)}$ box and examine the corresponding classical value and the optimal local deterministic winning strategies for this game. Later, we will demonstrate how to exclude nonlocal correlations on the boundary of the no-signaling correlation set, with a dimension of up to $4k-8$, provided the correlations can be decomposed as a convex combination of the $PR^{(k)}$ box and certain optimal local deterministic behaviors of the game $g_k$.

\begin{figure}[t]
		\centering
		\subfigure[The game $g_3$ and its corresponding no-signaling graph $G_3$.]{
			\begin{minipage}[t]{0.35\textwidth}
				\centering
				\includegraphics[width=1\textwidth]{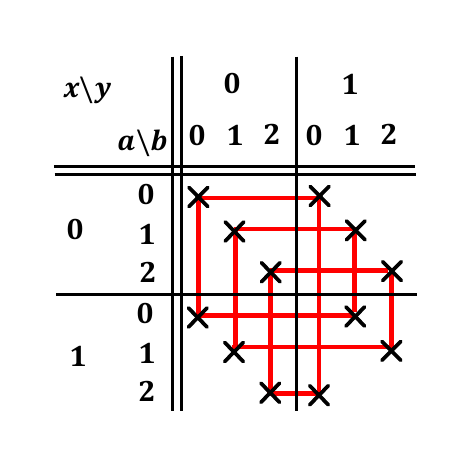}
			\end{minipage}
		}\quad \quad \quad
		\subfigure[One optimal local deterministic box for $g_3$ and its no-signaling graph]{
			\begin{minipage}[t]{0.35\textwidth}
				\centering
				\includegraphics[width=1\textwidth]{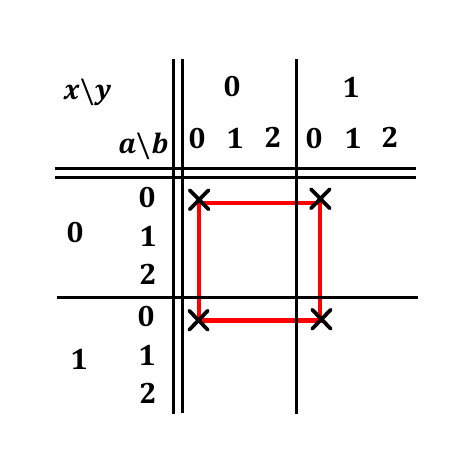}
			\end{minipage}
		}
		\caption{(a) $g_3$ is a non-local game in the $(2,2,3)$ Bell scenario with its input distribution being uniform and the winning conditions are labeled by \ding{53}. The no-signaling graph $G_3$ is represented with the vertices labeled by \ding{53} and the red lines denoting edges. (b) One local deterministic box that achieves the optimal classical value $\frac{3}{4}$ of the game $g_3$. The probabilities for the events labeled by \ding{53} are $1$ and others are $0$. The no-signaling graph of this local deterministic box is shown with vertices labeled by \ding{53} and the red lines denoting edges. One can see that the intersection of these two graphs is a path consisting of $3$ vertices and $2$ edges.} 
	\label{ns_graph}
\end{figure}

The nonlocal game $g_k$, which is the unique game defined by $PR^{(k)}$, is defined as follows: the input distribution is uniform, i.e., $\Pi(x,y)=1/4,\forall (x,y)$, and the winning condition is set to be $W(a,b,x,y)=1$ for all events $(a,b,x,y)$ for which $p(a,b|x,y)\neq 0$ in $PR^{(k)}$ and $W(a,b,x,y)=0$ for all other events. It's clear that the no-signaling value of this game $g_k$ is $1$, since this is achieved by the behavior $PR^{(k)}$. On the other hand, the classical value of the game $g_k$ is $3/4$, this can be seen as follows. The classical value is necessarily achieved by a local deterministic strategy (being the extreme points of the classical polytope) and for any local deterministic strategy $L_i^{(k)}$, the winning condition can be satisfied for at most three out of the four pairs of inputs $(x,y)$ (and in fact this is achievable by simply outputting $a = b = 0$ for $x, y \in \{0,1\}$).  
Thus, if $L_i^{(k)}$ is an optimal local deterministic strategy for the game $g_k$, it achieves the optimal classical value of $3/4$. We now proceed to show that there are $4k$ local deterministic boxes $L_i^{(k)}, i \in \{1, \ldots, 4k\}$, that achieve this classical value.
To see this we apply the notion of a {\em no-signaling graph} defined in~\cite{ramanathan2016no} to the game $g_k$, denoting the corresponding no-signaling graph by $G_{k}$. 
\begin{definition}[~\cite{ramanathan2016no}] For any non-local game $g$, we define the no-signaling graph $G_{NS}(g)=(V, E)$ associated with the game to have set of vertices $v \in V$, each of which is labeled by a set of inputs and outputs that wins the game, $v=\left(\bm{a}^{(v)}, \bm{x}^{(v)}\right)$. Two vertices $v, v^{\prime} \in V$ are connected be an edge if $\exists S \subseteq$ $[n]$ with $|S|=n-1$ such that $\left(a_i^{(v)}=a_i^{\left(v^{\prime}\right)} \wedge x_i^{(v)}=\right.$ $\left.x_i^{\left(v^{\prime}\right)}\right) \forall i \in S$.
\end{definition}
See the illustration of $g_3$ and one of its optimal local deterministic winning boxes, along with the no-signaling graph $G_{3}$ in Fig.~\ref{ns_graph}.

Note that the original definition of the no-signaling graph is associated with non-local games, but it naturally extends to  graphs $G_{l,i}$ representing local deterministic boxes $L_i^{(k)}$ in which the vertices $v$ of $G_{l,i}$ are labeled by the events $v=\left(\bm{a}, \bm{x}\right)$ for which $p\left(\bm{a}| \bm{x}\right)=1$ in $L_i^{(k)}$, and the edges are defined as before. 
Thus, suppose that $L_i^{(k)}$ is an optimal classical strategy of $g_k$, meaning that exactly three of the winning conditions are satisfied by $L_i^{(k)}$, then the intersection of their no-signaling graphs is directly seen to be a path consisting of $3$ vertices and $2$ edges (see the example of $k=3$ in Fig.~\ref{ns_graph}). Under this consideration, given a no-signaling graph $G_{l,i}$ of $L_i^{(k)}$ that is optimal for the game $g_k$, we can therefore define a unique 2-edge path in $G_{k}$. On the other hand, given any 2-edge path in $G_{k}$, we can also clearly construct a unique optimal local deterministic strategy for $g_k$. Thus there is a one-to-one correspondence between the optimal local deterministic strategies of $g_k$ and the 2-edge paths in $G_{k}$. On the other hand, the number of 2-edge paths of $G_{k}$ is the same as the number of vertices in $G_{k}$, which is $4k$.

Due to this structure, we can label each of these $4k$ local deterministic boxes by its event that is not in common with the no-signaling graph of $PR^{(k)}$, these are written as $L^{(k)}_{(a=b+1,b|x=0,y=0)},\forall b\in\{0,\ldots, k-1\}; L^{(k)}_{(a,b=a+1 |x=0,y=1)} \forall a\in\{0,\ldots, k-1\}; L^{(k)}_{(a,b=a+1 |x=1,y=0)}$ $ \forall a\in\{0,\ldots, k-1\}$ and $L^{(k)}_{(a,b=a|x=1,y=1)},\forall a\in\{0,\ldots, k-1\}$ (here we denote $k:=0$). 
Additionally, $PR^{(k)}$ is adjacent to these $4k$ local deterministic boxes in the $\mathrm{NS}(2,2,k)$ polytope.
Due to the structure of the no-signaling graph $G_k$ and these $4k$ optimal local deterministic strategies discussed above, the convex hull of $PR^{(k)}$ and any one of these $4k$ local deterministic boxes is equivalent under the relabeling of the input and output indices. Without loss of generality, we take the local deterministic box to be $L_{(a=0,b=0|x=1,y=1)}^{(k)}$. No other local deterministic box lies on the region formed by the convex hull of $PR^{(k)}$ and $L_{(a=0,b=0|x=1,y=1)}^{(k)}$, as any other local deterministic box contributes probabilities to events that have zero probability in both $PR^{(k)}$ and $L_{(a=0,b=0|x=1,y=1)}^{(k)}$. Additionally, no nonlocal extreme box lies on this region. If such a nonlocal extreme box $\{p(a,b|x,y)\}$ existed, then $p(a=0,b=0|x=1,y=1) \neq 0$, otherwise, it would coincide with $PR^{(k)}$. On the other hand, for any box that lies on this region, due to the definition of $PR^{(k)}$, the non-zero probability events for settings $x,y$ where $x \cdot y = 0$ must be correlated, i.e., $a = b$. These two conditions together contradict the fact that all nonlocal vertices in this scenario are of the form shown in Eq.~\eqref{eq_prk}.

We proceed to determine the quantum realization of a box $\mathrm{P}$ that belongs to the region formed by the convex hull of $PR^{(k)}$ and $4k-8$ of the $4k$ local deterministic boxes listed above. The eight excluded boxes are $L^{(k)}_{(a=0,b=k-1|x=0,y=0)},L^{(k)}_{(a=k-1,b=k-2|x=0,y=0)},$ $ L^{(k)}_{(a=k-2,b=k-1 |x=0,y=1)}$,$L^{(k)}_{(a=k-1,b=0 |x=0,y=1)}, L^{(k)}_{(a=k-2,b=k-1 |x=1,y=0)},L^{(k)}_{(a=k-1,b=0 |x=1,y=0)}$ and $L^{(k)}_{(a=0,b=0|x=1,y=1)}$, $L^{(k)}_{(a=k-1,b=k-1|x=1,y=1)}$. More specifically, we consider a box $\mathrm{P}$ that can be decomposed as follows:
\begin{equation}\label{decom_prk}
	\begin{split}
		\mathrm{P}=c_{NS}PR^{(k)} &+\sum_{b=0}^{k-3} c_{(a=b+1,b|0,0)}\cdot L^{(k)}_{(a=b+1,b|x=0,y=0)}+ \sum_{a=0}^{k-3} c_{(a,b=a+1 |x=0,y=1)}\cdot L^{(k)}_{(a,b=a+1 |x=0,y=1)}\\
		&+ \sum_{a=0}^{k-3} c_{(a,b=a+1 |x=1,y=0)}\cdot L^{(k)}_{(a,b=a+1 |x=1,y=0)}+\sum_{a=1}^{k-2} c_{(a,b=a |x=1,y=1)}\cdot L^{(k)}_{(a,b=a |x=1,y=1)}
	\end{split}
\end{equation}

where the coefficients $c$ are non-negative and their sum is equal to $1$ and $c_{NS}\neq 0$. We claim that $\mathrm{P}\notin \mathrm{Q}(2,2,k)$. This is because we find an almost quantum Bell inequality represented by $\widetilde{M}^{(k)}$ that rules out $\mathrm{P}$. 
\begin{equation}\label{m_k}
	\widetilde{M}^{(k)}=\left(
	\begin{array}{ccccc}
		4 & 4_{1\times (k-1)^2 } & 0 & 0_{1\times (k-1)^2 } & 1 \\ 
		4_{(k-1)^2\times 1 } & 8_{(k-1)^2\times (k-1)^2 } & 4_{(k-1)^2\times 1 } & 0_{(k-1)^2\times (k-1)^2 } & 0_{(k-1)^2\times 1 } \\ 
		0 & 4_{1\times (k-1)^2 } & 8 & 4_{1\times(k-1)^2 } & 0 \\ 
		0_{(k-1)^2\times 1 } & 0_{(k-1)^2\times (k-1)^2 } & 4_{1(k-1)^2\times 1 } & 8_{(k-1)^2\times (k-1)^2 } & 4_{(k-1)^2\times 1 } \\ 
		1 & 0_{1\times (k-1)^2 } & 0 & 4_{1\times (k-1)^2 } & 4
	\end{array}\right)
\end{equation}
where $4_{1 \times (k-1)^2}$ refers to a $1 \times (k-1)^2$ submatrix with all entries being $4$. This matrix Eq.~\eqref{m_k} can be seen as an extension of the matrix Eq.~\eqref{1bchain} which is used to exclude points on the 1-dimensional faces of $\mathrm{NS}(2,2,2)$ from $\mathrm{Q}(2,2,2)$. The matrix Eq.~\eqref{1bchain} has $\widetilde{M}_1\succ 0$, so that automatically we have $\widetilde{M}^{(k)}\succ 0$.

If we relabel the entries of $|\mathrm{P}\rangle$ which is the vector form of the box $\mathrm{P}$, such that the first $2(k-1)^2+3$ entries of it correspond to the probabilities of events $(k-1,k-1|0,1), (a,b|0,0)\forall a,b\in\{0,k-2\}, (k-1,k-1|0,0) , (a,b|1,0)\forall a,b\in\{0,k-2\}, (k-1,0|1,1)$ in order, we obtain the following inequality:
\begin{equation}
	\langle \mathrm{P}|\widetilde{M}^{(k)} |\mathrm{P}\rangle-\sum_{i=1}^{2(k-1)^2+3} \widetilde{M}_{(i,i)}^{(k)}|\mathrm{P}\>_i=2p(k-1,k-1|0,1)\cdot p(k-1,0|1,1)>0.
\end{equation} 
Thus all points $\mathrm{P}$ that can be written as Eq.~\eqref{decom_prk} are outside of $\mathrm{Q}(2,2,k)$.

Note that the correlations $\mathrm{P}$ considered in Eq.~\eqref{decom_prk} lie on the boundary of the no-signaling polytope $\mathrm{NS}(2,2,k)$ of dimension up to $4k-8$. On the one hand, the region formed by the convex hull of $PR^{(k)}$ and the $4k-8$ local deterministic boxes listed above is part of a face of $\mathrm{NS}(2,2,k)$. This follows from the definition of the no-signaling polytope, where the half-space representation is defined by normalization conditions and no-signaling conditions (both equality constraints), along with non-negativity conditions (inequality constraints). A face of a polytope is defined when some of these inequality constraints are saturated, here meaning that the probability of certain events becomes zero due to the saturation of non-negativity conditions. This is exactly the property that $\mathrm{P}$, as considered in Eq.~\eqref{decom_prk}, possesses.
Thus, while we cannot conclusively state that this convex hull forms a whole face of $\mathrm{NS}(2,2,k)$, we can still assert that it lies on some face of $\mathrm{NS}(2,2,k)$. 
On the other hand, in the convex combination of $PR^{(k)}$ and the $4k-8$ local deterministic boxes, each local deterministic box contributes uniquely to an event that no other box can. Therefore, these $4k-7$ boxes ($PR^{(k)}$ and the $4k-8$ local deterministic boxes) are linearly independent, and the dimension of the region formed by their convex hull is at most $4k-8$.

\section{Aumann's Agreement theorem and its quantum generalization}
    In the last section, we have seen how to use the quantum Bell inequality associated with the matrix $\widetilde{M}$ to identify whether a no-signaling box belongs to the almost-quantum set. This tool can be even applied to parameterized no-signaling boxes and by using this tool, we excluded several classes of no-signaling boxes from the almost-quantum set (and so the quantum correlation set). 
    In this section, we show that these theorems can be directly used as a crucial step in the proof of the quantum generalization of Aumann's Agreement theorem, which is recently defined in~\cite{contreras2021observers} as a candidate physical principle in the context of epistemics to pick out the quantum theory from general no-signaling ones. 

    Aumann's celebrated Agreement theorem states that `rational' agents cannot agree to disagree. More specifically, suppose a group of agents assign a probability to a given event based on their own partial information, and if the individual probabilities of each agent are common knowledge (or common certainty), then these individual probabilities must be the same, even though the probabilities are obtained by different Bayesian updates due to their different partial information. In plain language, the common knowledge (between two agents) means an infinite list of A knows B's probability (B knows A's probability), and B knows that A knows his probability (and also A knows that B knows her probability), and so on. This Agreement theorem is a basic requirement in classical epistemics and has been extensively used in decision theory and game theory.
    
    Recently, the quantum generalization of the Agreement theorem has been studied in ~\cite{contreras2021observers}.
    In particular, the authors establish two analogous notions of the Agreement theorem for the post-quantum (no-signaling) systems and characterize the no-signaling boxes that display these behaviors. And then they discuss the possibility of the quantum realizations of these no-signaling boxes, which concludes that the Agreement theorems still hold for observers of quantum systems.
    We list the main theorems of~\cite{contreras2021observers} in the following. In short, `disagreement' is the opposite concept of `agreement', i.e., the agents can agree to disagree.

    \begin{theorem}(Theorem 3 and Theorem 7 in~\cite{contreras2021observers})\\
		A two-input two-output no-signaling box gives rise to common
		certainty of disagreement if and only if it takes the form of Table~\ref{tab1}. \\
		A two-input two-output no-signaling box gives rise to singular disagreement if and only if it takes the form of Table~\ref{tab2}.
	\end{theorem}
    In definition, the singular disagreement means to replace the common certainty (common knowledge) requirement for a given event with the requirement that Alice's probability is 1 and Bob's probability is 0.

	\vspace{0.5em}
        \hspace{-0.85cm}
	\begin{minipage}{\textwidth}
		\begin{minipage}[t]{0.4\textwidth}
		\makeatletter\def\@captype{table}
		\resizebox{!}{1.08cm}{
			\begin{tabular}{|c|c|c|c|c|}
			\hline$x y \backslash a b$ & 00 & 01 & 10 & 11 \\
			\hline \hline 00 & $r$ & 0 & 0 & $1-r$ \\
			\hline 01 & $r-s$ & $s$ & $-r+t+s$ & $1-t-s$ \\
			\hline 10 & $t-u$ & $u$ & $r-t+u$ & $1-r-u$ \\
			\hline 11 & $t$ & 0 & 0 & $1-t$ \\
			\hline
		\end{tabular}}
		\caption{Parametrization of two-input two-output no-signaling boxes with common certainty of disagreement. Here, $r, s, t, u \in [0,1]$ are the values that ensure all entries of the box are non-negative, with $r > 0$ and $s - u \neq r - t$.}
		\label{tab1}
		\end{minipage}
		\quad
		\begin{minipage}[t]{0.58\textwidth}
		\makeatletter\def\@captype{table}
		\resizebox{!}{1.08cm}{
		\begin{tabular}{|c|c|c|c|c|}
		\hline$x y \backslash a b$ & 00 & 01 & 10 & 11 \\
		\hline \hline 00 & $s$ & $t$ & $1-s-u-t$ & $u$ \\
		\hline 01 & 0 & $s+t$ & $r$ & $1-s-t-r$ \\
		\hline 10 & $1-u-t$ & $u+t+r-1$ & 0 & $1-r$ \\
		\hline 11 & $r$ & 0 & 0 & $1-r$ \\
		\hline
		\end{tabular}}
		\caption{Parametrization of two-input two-output no-signaling boxes with singular disagreement. Here, $r, s, t, u \in [0,1]$ are the values that ensure all entries of the box are non-negative, with  $s>0$, and $s+t \neq 0$ and $u+t \neq 1$.}
		\label{tab2}
		\end{minipage}
	\end{minipage}

	\begin{theorem}(Theorem 4 and Theorem 8 in~\cite{contreras2021observers})
		No two-input two-output quantum box can give rise to common certainty of disagreement or singular disagreement.
	\end{theorem}
    The authors prove this theorem by deriving contradictions with Tsirelson's theorem and then use this theorem as the building blocks to extend to general scenarios with arbitrary numbers of agents, inputs and outputs.
    However, Tsirelson's theorem only holds for a subset of nonlocal games named the correlation Bell expressions (XOR games), in which only the correlators of the form  $\<A_xB_y\>$ appear but no local marginals $\<A_x\>$ or $\<B_y\>$. And by using Tsirelson's theorem, one can only get the quantum box that is achieved by the maximally entangled state with the local marginals being $\<A_x\>=0,\<B_y\>=0,\forall x,y.$ Thus these quantum boxes are not general and a general proof of this theorem is needed.

	\begin{theorem}\label{theo5}
		No two-input two-output quantum box or almost-quantum box can give rise to common certainty of disagreement or singular disagreement.
	\end{theorem}
    We present a general proof of this theorem by showing that these two tables are on the $4$-dimensional faces of the $(2,2,2)$ no-signaling polytope (we defer the details to Appendix.~\ref{Asec2}). Thus according to Theorem~\ref{theo1}, a direct consequence of it is that both tables are outside of the quantum correlation set and also the almost-quantum set. This result also answers the open question left in the discussion section of~\cite{contreras2021observers} that the Agreement still holds in the almost-quantum correlations. Almost-quantum correlations is one of the generalized quantum theories that is strictly larger than the quantum correlation set and satisfies almost all the reasonable physical and information-theoretic principles (except the principle of information causality, for which a proof is not known). 
    This result then adds one more common feature in the context of epistemics to quantum theory and almost-quantum correlation.

\section{Quantum Bell Inequalities and Self-testing}
In the previous sections, we proved that there are no quantum realizations for nonlocal boxes on the boundary of $\mathrm{NS}(2,2,k)$ of dimension $d \leq 4k-8$, with $k>2$. As stated earlier, the well-known Hardy paradox shows that quantum theory allows for a realization of a point on a five-dimensional face of the no-signaling set $\mathrm{NS}(2,2,2)$. In ~\cite{zhao2022tilted}, we extended this point to a whole region on the five-dimensional face that admits quantum realization (in an effort to find an optimal point for DI randomness extraction). Furthermore, we showed that these points allow to self-test all pure two-qubit entangled states except the maximally entangled state. 

In this section, we present a class of tight quantum Bell inequalities in the $(2,m,2)$ Bell scenario where each player performs $m\geq 2$ binary measurements. Let's denote by $A_i$ and $B_i\,(i=1,\ldots,m)$ the binary observable (outcomes $\pm1$) of Alice and Bob respectively. From the Tsirelson characterization of the set of quantum correlations in this scenario and by considering the positive semidefinite completion problem for the set of correlators in this cycle scenario~\cite{barrett1993real}, we obtain one boundary of the set of quantum correlations $\langle A_x B_y \rangle$ as follows.

\begin{proposition}
In the $(2,m,2)$ scenario, consider the $2m$ operators $\{A_1,\ldots,A_m,B_1,\ldots,B_m\}$ with binary outcomes $\pm1$ and assumed to fulfill $[A_x,B_y]=0$ and an unknown state $|\psi\>$. Define the unit vectors $\vec{A}_x=A_x|\psi\rangle$ and $\vec{B}_y=B_y|\psi\rangle$, and the observed correlations $E_{x,y}\equiv \<\psi|A_xB_y|\psi\>=\vec{A}_x^{\dag}\vec{B}_y=\cos\alpha_{x,y}$, where $\alpha_{x,y}\geq 0$ is the angle between  $\vec{A}_x$ and $\vec{B}_y$.
The set of correlations $E_{x,y}$ achievable in quantum theory (up to relabeling the index of operators) has the boundary:
\begin{equation}\label{boundary}
	\sum\limits_{i=1}^{m-1} \left(\alpha_{i,i}+\alpha_{i+1,i}\right)=\alpha_{1,m}-\alpha_{m,m}
\end{equation}
\end{proposition}

If this set of correlations $E_{x,y}$ are available in quantum theory, then one can construct a Gram matrix by using these unit vectors $\vec{A}_x, \vec{B}_y$ such that the corresponding entries of the Gram matrix are equal to the value of correlations $E_{x,y}$. The real positive definite matrix completion problem for simple cycle graphs is studied in~\cite{barrett1993real}, where the vertices of the cycle graphs are defined by the entries of the real positive definite matrix and two vertices are connected by an edge if and only if the corresponding entries of the matrix are in the same column or in the same row. Thus, by listing the vectors $\vec{A}_x, \vec{B}_y$ as the column of $V_{gram}$ in the order that $V_{gram}=[\vec{A}_1, \vec{B}_1, \vec{A}_2, \vec{B}_2,\ldots,\vec{A}_m, \vec{B}_m ]$ and defining the Gram matrix as $M_{gram}:=V_{gram}^{T}V_{gram}$, the solution of this Gram matrix completion problem then suggests the above result Eq.~\eqref{boundary}. However, the semidefinite completion problem for the cycle only considers correlations of the form $E_{x,y}$ for $y = x, x+1$, and as such only gives an outer approximation boundary to the quantum set. In the following, we prove the self-testing nature of the correlations satisfying Eq.~\eqref{boundary} and in the process show that the boundary is tight for the set of quantum correlations (excluding the marginals).

This statement generalizes the well-known boundary of the set of quantum correlations in the $(2,2,2)$ scenario characterized by  Tsirelson-Landau-Masanes as \cite{cirel1980quantum,landau1988empirical,masanes2003necessary}:
\begin{equation}
\sum_{(x,y) \neq (i,j)} \arcsin{\left(\langle A_{x} B_y \rangle \right)} - \arcsin{\left(\langle A_{i} B_j \rangle \right)} = \xi \pi,
\end{equation}
where $i, j \in \{1,2\}$ and $\xi = \pm 1$. 
One can also see that the corresponding points on the $\mathrm{Q}(2,m,2)$ boundary optimally violate a weighted Braunstein-Caves chained inequality expression of the form
\begin{equation}\label{w_chain}
I_{ch}^m:= \sum\limits_{i=1}^{m-1} \left(c_{i,i}\<A_iB_i\>+c_{i+1,i}\<A_{i+1}B_i\>\right) +c_{m,m} \<A_mB_m\>-c_{1,m}\<A_{1}B_m\>
\end{equation}
with appropriate coefficients $c_{i,j}$. We now proceed to show that points on this boundary serve to self-test the singlet state (along with suitable reference measurements). This generalizes the characterization of singlet self-testing boundary points in the $(2,2,2)$ scenario obtained in \cite{wang2016all} and the self-testing of the chain inequality proven in \cite{vsupic2016self} (where a robust result was derived).

\begin{figure}[htbp]
\centering
\includegraphics[width=0.6\textwidth]{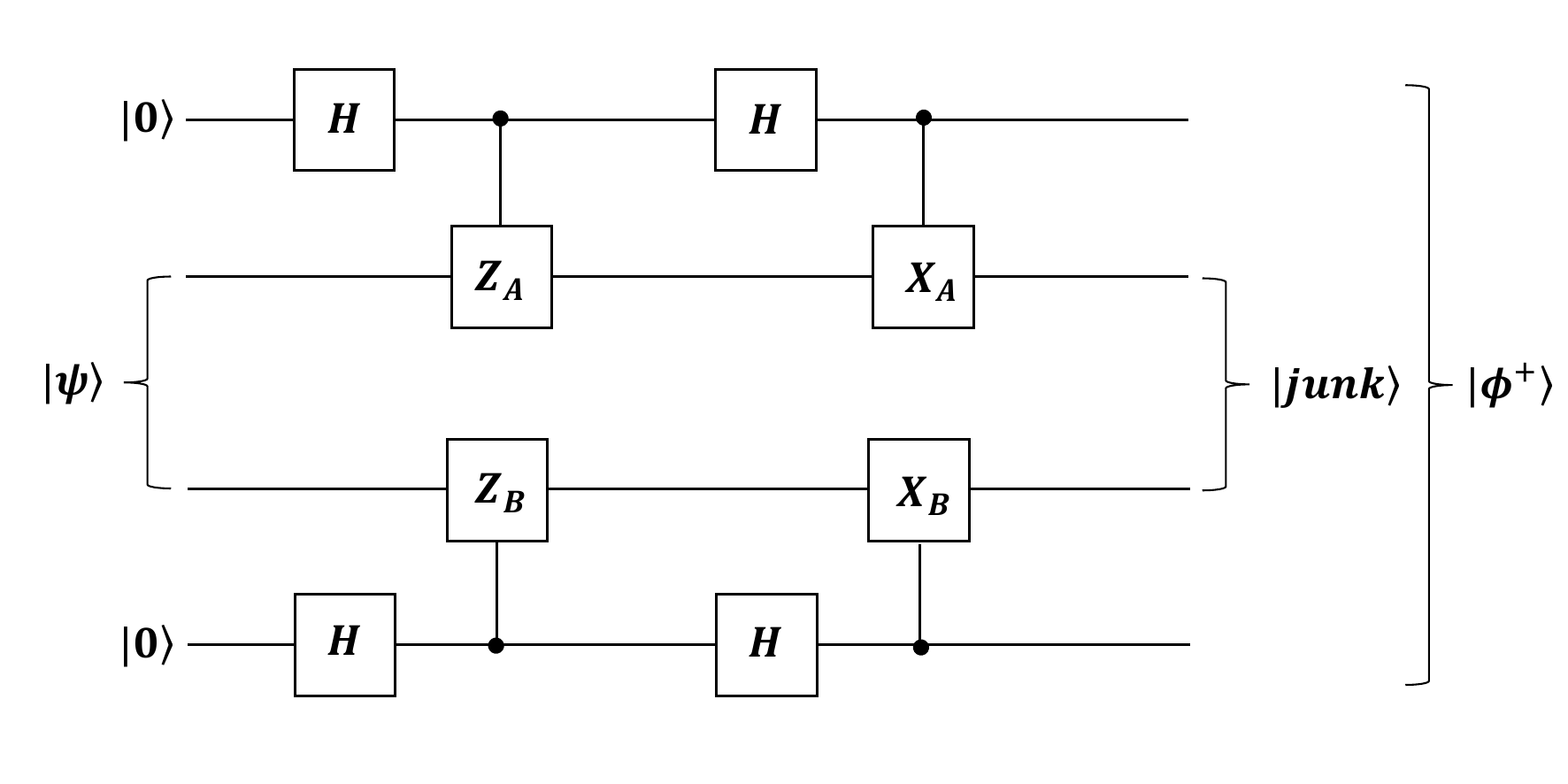}\\
\caption{Local isometry $\Phi$ used to self-test the singlet. $H$ is the Hadamard gate.}
\label{circuit}
\end{figure}
It's easy to verify that the singlet is self-tested by this circuit shown in Fig.~\ref{circuit} if these control operators are unitary (if they are not unitary, one can perform a regularization on the non-unitary operators) and satisfy \cite{mckague2012robust}:
\begin{equation}\label{condition_self_test}
\begin{split}
	Z_A |\psi\rangle &=Z_B |\psi\rangle \\
	X_A |\psi\rangle &=X_B |\psi\rangle \\
	X_A  Z_A |\psi\rangle &=-Z_A  X_A |\psi\rangle \\
	X_B  Z_B |\psi\rangle &=-Z_B  X_B |\psi\rangle 
\end{split}
\end{equation}
We will see how to construct these control operators $Z_{A/B}, X_{A/B}$ from the binary measurements $A_x,B_y$ in the weighted chain inequality Eq.~\eqref{w_chain}.

\begin{figure}[htbp]
\centering
\includegraphics[width=0.6\textwidth]{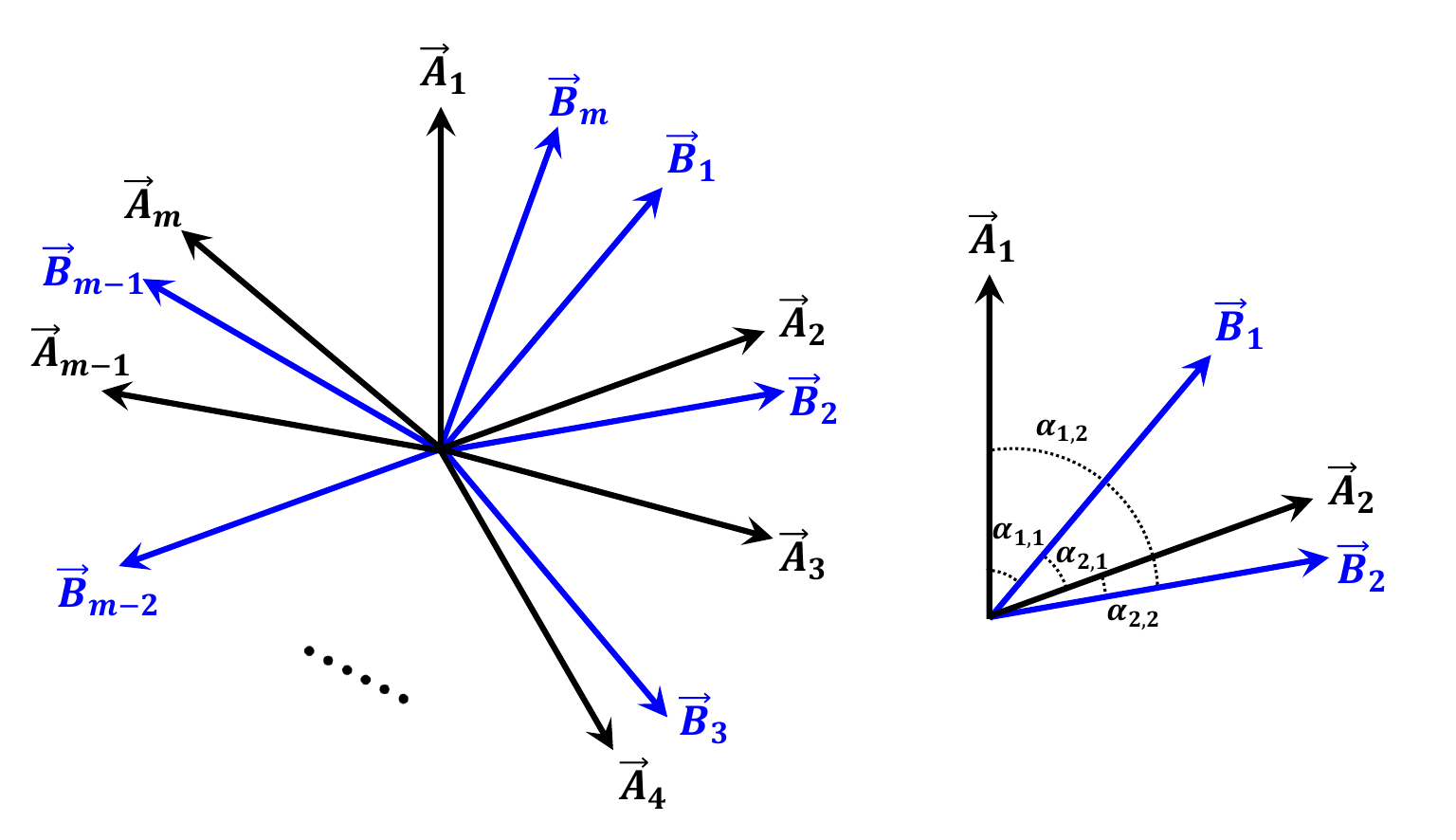}\\
\caption{All vectors $\vec{A}_i,\vec{B}_i\,\forall i=1,\ldots, m$  lie on the same plane. And $\alpha_{1,1}+\alpha_{2,1}+\alpha_{2,2}=\alpha_{1,2}$.}
\label{angle}
\end{figure}

Let's denote by $\theta_{i,j}(>0)$ the angles between $\vec{A}_i$ and $\vec{A}_j$. Given the scalar products of $\vec{A}_i$, $\vec{A}_{i+1}$ with $\vec{B}_i$, the angle $\theta_{i,i+1},\forall i=1,\ldots m-1$ must satisfy:
\begin{equation}
\begin{split}\label{up}
|\alpha_{i,i}-\alpha_{i+1,i}|\leq \theta_{i,i+1}\leq \alpha_{i,i}+\alpha_{i+1,i}.
\end{split}
\end{equation}
The lower and upper bounds are reached when $\vec{B}_i$ lies in the same plane of $\vec{A}_i$, $\vec{A}_{i+1}$.
And the angle $\theta_{m,1}$ satisfies:
\begin{equation}\label{low}
|\alpha_{m,m}-\alpha_{1,m}|\leq \theta_{m,1}\leq \alpha_{m,m}+\alpha_{1,m}
\end{equation}
Together with the boundary condition Eq.~\eqref{boundary}, this implies that all vectors $\vec{A}_i,\vec{B}_i\,\forall i=1,\ldots, m$  lie on the same plane, inequalities in Eq.~\eqref{up} reach the upper bound and inequalities in Eq.~\eqref{low} reach the lower bound. See Fig.~\ref{angle}.

In particular, we have 
\begin{equation}
\begin{split}
	\vec{B}_1=\frac{\sin \left(\alpha_{1,1}\right) \vec{A}_2+\sin \left(\alpha_{2,1}\right) \vec{A}_1}{\sin \left(\alpha_{1,1}+\alpha_{2,1}\right)} \\
	\vec{A}_2=\frac{\sin \left(\alpha_{2,2}\right) \vec{B}_1+\sin \left(\alpha_{2,1}\right) \vec{B}_2}{\sin \left(\alpha_{2,1}+\alpha_{2,2}\right)} .     
\end{split}
\end{equation}

Using the commutativity $[A_x,B_y]=0$, and the fact that $A_x^2=B_y^2=I$ (the outcomes of the operators are binary outcomes $\pm 1$), we obtain:
\begin{equation}
\begin{split}       
	\left(A_1 A_2+A_2 A_1\right)|\psi\rangle=2 \cos \left(\alpha_{1,1}+\alpha_{2,1}\right)|\psi\rangle\\
	\left(B_1 B_2+B_2 B_1\right)|\psi\rangle=2 \cos \left(\alpha_{2,2}+\alpha_{2,1}\right)|\psi\rangle
\end{split}
\end{equation}
Now, we can construct the control operators as:
\begin{equation}
\begin{split}
	Z_A  & =A_1 \\
	X_A  & =\frac{A_2-\cos \left(\alpha_{1,1}+\alpha_{2,1}\right) A_1}{\sin \left(\alpha_{1,1}+\alpha_{2,1}\right)} \\
	Z_B  & =\frac{\sin \left(\alpha_{1,2}\right) B_1-\sin \left(\alpha_{1,1}\right) B_2}{\sin \left(\alpha_{1,2}-\alpha_{1,1}\right)} \\
	X_B  & =\frac{\cos \left(\alpha_{1,1}\right) B_2-\cos \left(\alpha_{1,2}\right) B_1}{\sin \left(\alpha_{1,2}-\alpha_{1,1}\right)}
\end{split}
\end{equation}
We now verify these control operators satisfy the conditions listed in Eq.~\eqref{condition_self_test}.
\begin{equation}\label{self_zx}
\begin{split}
	\left(Z_AX_A+X_AZ_A\right)|\psi\>&=\frac{A_1A_2+A_2A_1-2\cos \left(\alpha_{1,1}+\alpha_{2,1}\right)}{\sin \left(\alpha_{1,1}+\alpha_{2,1}\right)}|\psi\>=0\\
	\left(Z_BX_B+X_BZ_B\right)|\psi\>&=\frac{\sin \left(\alpha_{1,2}+\alpha_{1,1}\right)(B_1B_2+B_2B_1)-\left[\sin \left(2\alpha_{1,1}\right)+\sin \left(2\alpha_{1,2}\right) \right]}{\sin^2 \left(\alpha_{1,2}-\alpha_{1,1}\right)}|\psi\>\\
	&=\frac{\sin \left(\alpha_{1,2}+\alpha_{1,1}\right)\left[B_1B_2+B_2B_1-2\cos \left(\alpha_{1,2}-\alpha_{1,1}\right) \right]}{\sin^2 \left(\alpha_{1,2}-\alpha_{1,1}\right)}|\psi\>\\
	&= \frac{2\sin \left(\alpha_{1,2}+\alpha_{1,1}\right)\left[\cos \left(\alpha_{2,2}+\alpha_{2,1}\right)-\cos \left(\alpha_{1,2}-\alpha_{1,1}\right) \right]}{\sin^2 \left(\alpha_{1,2}-\alpha_{1,1}\right)}|\psi\>=0\\  
\end{split}
\end{equation}
Note that we use $\sin \left(2\alpha_{1,1}\right)+\sin \left(2\alpha_{1,2}\right)=2\sin(\alpha_{1,1}+\alpha_{1,2})\cos(\alpha_{1,1}-\alpha_{1,2})$ to simplify above calculation. And we also have:
\begin{equation}\label{self_zz1}
\begin{split}
	\<\psi| Z_AZ_B|\psi\>&=\<\psi| \frac{Z_AZ_B+Z_BZ_A}{2}|\psi\>\\
 &=\<\psi|\frac{\sin \left(\alpha_{1,2}\right) (A_1B_1+B_1A_1)-\sin \left(\alpha_{1,1}\right) (A_1B_2+B_2A_1)}{2\sin \left(\alpha_{1,2}-\alpha_{1,1}\right)}|\psi\>\\
\end{split}
\end{equation}

From Fig.~\ref{angle}, we have $\vec{A}_1=\frac{\sin \left(\alpha_{1,2}\right) \vec{B}_1+\sin \left(\alpha_{1,1}\right) \vec{B}_2}{\sin \left(\alpha_{1,1}+\alpha_{1,2}\right)}$, thus we simplify the above Eq.~\eqref{self_zz1} to

\begin{equation}\label{self_zz2}
\begin{split}
	\<\psi| Z_AZ_B|\psi\>&=\<\psi| \frac{2\sin^2(\alpha_{1,2})-2\sin^2(\alpha_{1,1})}{2\sin \left(\alpha_{1,2}-\alpha_{1,1}\right)\sin \left(\alpha_{1,1}+\alpha_{1,2}\right)}|\psi\>\\
 &=\<\psi| \frac{\cos(2\alpha_{1,1})-\cos(2\alpha_{1,2})}{2\sin \left(\alpha_{1,2}-\alpha_{1,1}\right)\sin \left(\alpha_{1,1}+\alpha_{1,2}\right)}|\psi\>=1.
\end{split}
\end{equation}
Note that we use the equality $\cos \left(2\alpha_{1,1}\right)-\cos \left(2\alpha_{1,2}\right)=-2\sin(\alpha_{1,1}+\alpha_{1,2})\sin(\alpha_{1,1}-\alpha_{1,2})$ in above calculation. And similarly, $\<\psi| X_AX_B|\psi\>=1$ holds. Together with Eq.~\eqref{self_zx},~~\eqref{self_zz2} this implies that the self-testing conditions Eq.~\eqref{condition_self_test} hold.

Any pair of vectors $\vec{A}_i,\vec{B}_j$ can be written as 
\begin{equation}
\begin{split}
	\vec{A}_i= a_{z,i}Z_A|\psi\>+a_{x,i}X_A|\psi\>\\
	\vec{B}_j= b_{z,j}Z_B|\psi\>+b_{x,j}X_B|\psi\>
\end{split}
\end{equation}
where $a_{z,i}=\vec{A}_i\cdot Z_A|\psi\>,a_{x,i}=\vec{A}_i\cdot X_A|\psi\>,b_{z,j}= \vec{B}_j\cdot Z_B|\psi\>, b_{x,j}= \vec{B}_j\cdot X_B|\psi\> $.
Thus when the isometry is applied to any pair of operators, we get
\begin{equation}\label{selftest_pair}
\begin{split}
	\Phi\left(A_i  B_j \left|\psi \right\rangle|00\rangle\right) &=  a_{z,i} b_{z,j}  \Phi\left(Z_A  Z_B \left|\psi \right\rangle|00\rangle\right)+a_{z,i} b_{x,j}  \Phi\left(Z_A  X_B \left|\psi \right\rangle|00\rangle\right) \\
	& +a_{x,i}b_{z,j}  \Phi\left(X_A  Z_B \left|\psi \right\rangle|00\rangle\right)+a_{x,i} b_{x,j}  \Phi\left(X_A  X_B \left|\psi \right\rangle|00\rangle\right)\\
	&=\left(a_{z,i} b_{z,j}+ a_{x,i} b_{x,j}\right) \Phi\left( \left|\psi \right\rangle|00\rangle\right) + \left(a_{x,i}b_{z,j}-a_{z,i} b_{x,j}\right) \Phi\left(X_A  Z_B \left|\psi \right\rangle|00\rangle\right) 
\end{split}
\end{equation}
And the first term in Eq.~\eqref{selftest_pair} is 
\begin{equation}
\begin{split}
	\Phi\left( \left|\psi \right\rangle|00\rangle\right)&=\frac{1}{4}\left(\left(\mathbb{I}+Z_A \right)\left(\mathbb{I}+Z_B\right)\left|\psi\right\rangle|00\rangle+X_A \left(\mathbb{I}-Z_A \right)\left(\mathbb{I}+ Z_B\right)\left|\psi \right\rangle|10\rangle\right. \\
	&\left.+X_B\left(\mathbb{I}+Z_A \right)\left(\mathbb{I}- Z_B\right)\left|\psi \right\rangle|01\rangle+X_A  X_B\left(\mathbb{I}-Z_A \right)\left(\mathbb{I}- Z_B\right)\left|\psi \right\rangle|11\rangle\right)\\
	&=\frac{1}{2}\left(\left(\mathbb{I}+Z_A \right)\left|\psi\right\rangle|00\rangle+X_A  X_B\left(\mathbb{I}-Z_A \right)\left|\psi \right\rangle|11\rangle+0\cdot \left|\psi \right\rangle|10\rangle+0\cdot \left|\psi \right\rangle|01\rangle \right)\\
	&=\frac{1}{\sqrt{2}}\left(\mathbb{I}+Z_A \right)|\psi\>\frac{1}{\sqrt{2}}\left(|00\rangle+|11\rangle\right)=|junk\>|\phi^+\>\\
\end{split}
\end{equation}
The second term in Eq.~\eqref{selftest_pair} is 
\begin{equation}
\begin{split}
	&\Phi\left(X_A  Z_B \left|\psi \right\rangle|00\rangle\right) =\frac{1}{4}(\left(\mathbb{I}+Z_A \right)\left(\mathbb{I}+ Z_B\right) X_A  Z_B \left|\psi \right\rangle|00\rangle+X_A \left(\mathbb{I}-Z_A \right)\left(\mathbb{I}+ Z_B\right) X_A  Z_B \left|\psi \right\rangle|10\rangle \\
	& +X_B\left(\mathbb{I}+Z_A \right)\left(\mathbb{I}- Z_B\right) X_A  Z_B \left|\psi \right\rangle|01\rangle +X_A  X_B\left(\mathbb{I}-Z_A \right)\left(\mathbb{I}- Z_B\right) X_A  Z_B \left|\psi \right\rangle|11\rangle)\\
	&=\frac{1}{4}(X_A\left(\mathbb{I}-Z_A \right)\left(\mathbb{I}+ Z_B\right) Z_B \left|\psi \right\rangle|00\rangle+\left(\mathbb{I}+Z_A \right)\left(\mathbb{I}+ Z_B\right)   Z_B \left|\psi \right\rangle|10\rangle \\
	& +\left(\mathbb{I}+Z_A \right)\left(\mathbb{I}+ Z_B\right) X_A  X_B Z_B \left|\psi \right\rangle|01\rangle +  X_B\left(\mathbb{I}+Z_A \right)\left(\mathbb{I}- Z_B\right)  Z_B \left|\psi \right\rangle|11\rangle)\\
	&=\frac{1}{4} (\left(\mathbb{I}+Z_A \right)\left(\mathbb{I}+ Z_B\right) \left|\psi \right\rangle|10\rangle+\left(\mathbb{I}+Z_A \right)\left(\mathbb{I}+ Z_B\right) X_A  X_B Z_B \left|\psi \right\rangle|01\rangle+0\cdot \left|\psi \right\rangle|00\rangle+0\cdot \left|\psi \right\rangle|11\rangle)\\
	&=\frac{1}{\sqrt{2}}\left(\mathbb{I}+Z_A \right)\left|\psi \right\rangle\frac{1}{\sqrt{2}}\left(|10\rangle-|01\rangle\right)=|junk\>\sigma_{x,A}\sigma_{z,B}|\phi^+\>\\
\end{split}
\end{equation}
So Eq.~\eqref{selftest_pair} can be written as:
\begin{equation}
\begin{split}
	&\Phi\left(A_i  B_j \left|\psi \right\rangle|00\rangle\right) =\left(a_{z,i} b_{z,j}+ a_{x,i} b_{x,j}\right) \Phi\left( \left|\psi \right\rangle|00\rangle\right) + \left(a_{x,i}b_{z,j}-a_{z,i} b_{x,j}\right) \Phi\left(X_A  Z_B \left|\psi \right\rangle|00\rangle\right) \\
	&=\left(a_{z,i} b_{z,j}+ a_{x,i} b_{x,j}\right)|junk\>|\phi^+\> + \left(a_{x,i}b_{z,j}-a_{z,i} b_{x,j}\right)|junk\>\sigma_{x,A}\sigma_{z,B}|\phi^+\>\\
	&=|junk\> \left (a_{z,i}\sigma_{z,A} b_{z,j}\sigma_{z,B}+a_{x,i}\sigma_{x,A} b_{x,j}\sigma_{x,B}+a_{x,i}\sigma_{x,A} b_{z,j} \sigma_{z,B}+a_{z,i}\sigma_{z,A} b_{x,j}\sigma_{x,B}\right)|\phi^+\>\\
	&=|junk\> \widetilde{A}_i\widetilde{B}_j|\phi^+\>
\end{split}
\end{equation}
Note that $\sigma_{x,A}\sigma_{x,B}|\phi^+\>=\sigma_{z,A}\sigma_{z,B}|\phi^+\>=|\phi^+\>$.
Similarly for the local operators \\$\Phi\left(A_i\left|\psi \right\rangle|00\rangle\right)=|junk\> \widetilde{A}_i|\phi^+\>$ and $\Phi\left(B_j\left|\psi \right\rangle|00\rangle\right)=|junk\> \widetilde{B}_j|\phi^+\>$, where $\widetilde{A}_i, \widetilde{B}_j$ are the reference measurements.

\section{Common faces of the Quantum and Classical Correlation sets}
In the previous sections, we have studied the boundary of the quantum set with specific regard to its relation with the no-signaling boundary and to self-testing applications. In this section, we explore the region of the boundary of the quantum set that also serves as the boundary of the classical set. 
Motivated by finding the nonlocal games that have no quantum advantage, Linden et al. in \cite{linden2007quantum} first discovered a family of two-party inequalities that do not admit any quantum violation. These inequalities define the NLC (no advantage in nonlocal computation) information-theoretic principle.
As such, these regions serve as testing grounds and pointers towards information-theoretic means of constraining general non-signaling correlations and picking out quantum theory from among non-signaling theories. It was proven in~\cite{ramanathan2017tightness, linden2007quantum, ramanathan2021violation} that NLC games do not define facets of the classical correlation set. In~\cite{escola2020all, ramanathan2021violation}, the authors studied the geometry of the common faces of the quantum and classical correlation set and found that none of them are tight.

Specifically, we focus on the set of correlations alone, i.e., the correlators $\langle A_x B_y \rangle$ for binary observables $A_x, B_y$ excluding the local marginals $\langle A_x \rangle, \langle B_y \rangle$. Such correlation Bell inequalities are also termed XOR games, and we present systematic constructions of nontrivial XOR games with $m_a, m_b$ inputs for Alice and Bob respectively such that the quantum value of the game equals the classical value.

In this bipartite $m_a\times m_b$-inputs, $2\times 2$-outputs Bell scenario, we can represent the extreme points of the local set as a column vector with $2^{m_a+m_b}$ rows with all entries equal to either plus or minus one. Using the well-known Tsirelson characterization \cite{cirel1980quantum,Tsirel1987} of the set of quantum correlations in this scenario as an elliptope, and leveraging results on the facial structures of the set of correlation matrices \cite{laurent1996facial}, we obtain the following statement.

\begin{theorem}[\cite{laurent1996facial}]
\label{thm:Laurent}
If there exists a subset $R=\{v_1,...,v_{|R|}\}\subseteq L$, with cardinality $r=|R|\leq \log_2(m_a+m_b)$, and for any subset $I\subseteq\{1,...,|R|\}$, we have
\[\odot_{j\in\{1,...,r\}} v_{I,j}\neq 0\]
where 
\[v_{I,j}=\begin{cases}
	I_{2^{m_a+m_b}} + v_j &\text{if $j\in I$}\\
	I_{2^{m_a+m_b}} - v_j &\text{if $j\not\in I$}\\
\end{cases}\]
and $I_{2^{m_a+m_b}}$ is a column vector with $2^{m_a+m_b}$ rows and all entries equal to one and $\odot$ is the Hadamard product. Then $R$ forms the boundary of the local set which the quantum set saturates to. In such a case, the subset $R=\{v_1,...,v_{|R|}\}$ is said to be in general position.  
\end{theorem}

We utilize Theorem \ref{thm:Laurent} to construct a class of low-dimensional faces of the set of classical correlations that also serves as the boundary of the quantum set. In this respect, we recover the class of games discovered by Linden et al. in \cite{linden2007quantum} for the special case that $m_a = m_b = 2^k$ for $k \geq 2$. To do that, we first show that the set of vectors satisfying the conditions in Theorem~\ref{thm:Laurent} are unique in the following sense:

\begin{lemma}{\label{lem_unique}}
Let $m_a,m_b\in\mathbb{N}$ and $\mathbb{N}\ni r\leq \log_2(m_a+m_b)$. Let $\{v_1,...,v_r\}$ be a set of general positions vectors of $2^{m_a+m_b}$ rows. Then up to swapping rows and re-indexing, the first $2^r$ rows of them are unique. In particular, the unique choice for the first $2^r$ rows is the lexicographical ordering.
\end{lemma}
We defer the proof of this lemma to Appendix.~\ref{Asec4}. By this lemma, if $m_a=m_b=2^k$, where $k\geq 2$, $r=\log_2(m_a+m_b)=k+1$ and $\{v_1,...,v_r\}$ are in general positions, then $\{v_1,...,v_r\}$ are unique up to swapping rows and re-indexing. We now focus on this case.
For convenience reason, we first concatenate $v_1,...,v_r$ into one matrix, denoted by $G_{2^k}$ ( $G_{2^k}$ is a $2^{k+1}$ by $r$ matrix). We then define the $2^k$ by $2^k$ game matrix $\G_{2^k}$ as follows. The $(i,j)$ position of the game matrix $\G_{2^k}$ associates with Alice's $i^{th}$ input and Bob's $j^{th}$ input. In other words, the $(i,j)$ position of $\G_{2^k}$ associates with the $i^{th}$ row and the ${(2^k+j)}^{th}$ row of $G_{2^k}$. Let $r_i=(x_1,...,x_r)$ and $r_j=(y_1,...,y_r)$ be the $i^{th}$ row and the $(2^k+j)^{th}$ row of $G_{2^k}$ respectively. We define the $(i,j)$ entry of $\G_{2^k}$ by using an operator $\star$ on the $i^{th}$ row and the $(2^k+j)^{th}$ row of $G_{2^k}$ as follows:
$$
(\G_{2^k})_{i,j}=r_i\star r_j=\begin{cases}
1 &\text{if $x_1y_1+...+x_ry_r>0$}\\
-1 &\text{if $x_1y_1+...+x_ry_r<0$}\\
0 &\text{if $x_1y_1+...+x_ry_r=0$}\\
\end{cases}
$$

Under this construction, the game matrix $\G_{2^k}$ with $k\geq 4$ is of a specific form:

\begin{lemma}{\label{lem_main}}
Consider the game matrix $\G_{2^k}$, where $k\geq 4$ and denote $d=dim(\G_{2^k})=2^k$. The game matrix $\G_{2^k}$ has form
$$
\G_{2^k}=\begin{pmatrix}
	A^{(d)}_{2^{k-2}}&\G_{2^{k-2}}&\G_{2^{k-2}}&B^{(d)}_{2^{k-2}}\\
	\G_{2^{k-2}}&A^{(d)}_{2^{k-2}}&B^{(d)}_{2^{k-2}}&\G_{2^{k-2}}\\
	\G_{2^{k-2}}&B^{(d)}_{2^{k-2}}&A^{(d)}_{2^{k-2}}&\G_{2^{k-2}}\\
	B^{(d)}_{2^{k-2}}&\G_{2^{k-2}}&\G_{2^{k-2}}&A^{(d)}_{2^{k-2}}
\end{pmatrix}
$$
where $A^{(d)}_{2^{k-2}}=\begin{pmatrix}A^{(d)}_{2^{k-3}}&\G_{\frac{d}{4},2^{k-3}}\\\G_{\frac{d}{4},2^{k-3}}&A^{(d)}_{2^{k-3}}\end{pmatrix}$, $B^{(d)}_{2^{k-2}}=\begin{pmatrix}\overline{\G_{\frac{d}{4},2^{k-3}}}&B^{(d)}_{2^{k-3}}\\B^{(d)}_{2^{k-3}}&\overline{\G_{\frac{d}{4},2^{k-3}}}\end{pmatrix}$, $A^{(d)}_1=1$ and $B^{(d)}_1=-1$. 

$\G_{2^k,x}$ is defined to be the top left corner square block of $\G_{2^k}$ with dimension $x$ by $x$ and  $\overline{\G_{2^k,x}}$ is the top-most right-most corner square block of $\G_{2^k}$ of dimension $x$ by $x$.
\end{lemma}

This lemma and two concrete examples for $k=2$ and $k=3$ allow us to prove that for any $k\geq 2$, the game matrix $\G_{2^k}$ is diagonal in the Hadamard basis (details are in Appendix.~\ref{Asec4}), which reproduces the class of games discovered by Linden et al. in \cite{linden2007quantum}.

\section{Conclusions and Open problems}

The boundary of the set of quantum behaviors is notoriously difficult to characterise and yet of fundamental and practical interest for device-independent applications. In this work, we investigated the boundary by means of studying the class of almost-quantum Bell inequalities. We proved the utility of this class of inequalities in ruling out a large portion of post-quantum no-signaling behaviors, extending previous results obtained through nonlocality distillation and the collapse of communication complexity. As a direct consequence of this investigation, we prove that Aumann's Agreement theorem holds for the quantum systems as well as the almost-quantum correlations, the investigation of the utility of this principle in picking out quantum theory is of great interest for the future. In the scenario of two players with $m$ binary measurements, we provide a tight set of quantum Bell inequalities and prove the self-testing property of the correlations therein. Finally, we utilised the almost quantum Bell inequalities to derive a general form of the principle of no advantage in nonlocal computation.

The following are few of the possible directions for extending this work:

1. 
In the future, it would be very interesting to derive a class of quantum Bell inequalities from SDP approximations of the quantum set going beyond the almost-quantum class. 
2. Quantum realizations of the no-signaling boundary being crucial for device-independent randomness amplification, our study of the experimentally friendly two-input inequalities serves to pick out the optimal quantum correlations suited for this task. 
3. The Braunstein-Caves chained Bell inequality has found applications in many device-independent protocols, specifically in the scenario of achieving security against non-signaling adversaries.  We have explored the boundary of the quantum correlation set in this scenario and demonstrated their usefulness in self-testing the two-qubit singlet along with appropriate measurements. In the future, it would be interesting to see if such weighted chain inequalities achieve better performance in specific DI tasks \cite{ramanathan2018practical,ramanathan2021no}.

\section*{Acknowledgments}
The authors thank Pawe{\l} Horodecki for useful discussions. The authors acknowledge support from the Early Career Scheme (ECS) grant "Device-Independent Random Number Generation and Quantum Key Distribution with Weak Random Seeds" (Grant No. 27210620), the General Research Fund (GRF) grant "Semi-device-independent cryptographic applications of a single trusted quantum system" (Grant No. 17211122) and the Research Impact Fund (RIF) "Trustworthy quantum gadgets for secure online communication" (Grant No. R7035-21).


\appendix
\setcounter{table}{0}
\renewcommand{\thetable}{A\arabic{table}}
\setcounter{figure}{0}
\renewcommand{\thefigure}{A\arabic{figure}}
\setcounter{definition}{0}
\renewcommand{\thedefinition}{A\arabic{definition}}
\setcounter{theorem}{0}
\renewcommand{\thetheorem}{A\arabic{theorem}}
\setcounter{lemma}{0}
\renewcommand{\thelemma}{A\arabic{lemma}}
\setcounter{proposition}{0}
\renewcommand{\theproposition}{A\arabic{proposition}}
\setcounter{examp}{0}
\renewcommand{\theexamp}{A\arabic{examp}}
\setcounter{corollary}{0}
\renewcommand{\thecorollary}{A\arabic{corollary}}

\section{Excluding Nonlocal no-signaling Boxes from $\mathrm{Q}(2,2,2)$}\label{Asec1}
 
 \subsection*{Proof of Theorem~\ref{theo1}}
	\begin{theorem} \label{Atheo1}
		In the $(2,2,2)$ Bell scenario, let $\mathrm{P}$ be a point on a face of the no-signaling polytope $\mathrm{NS}(2,2,2)$ of dimension $d\leq 4$, such that $\mathrm{P}\notin \mathrm{L}(2,2,2)$, then $\mathrm{P}\notin \mathrm{Q}(2,2,2)$.
	\end{theorem}
	
\begin{proof}
In the main text, we have provided the proof to exclude the point $\mathrm{P}$ on the faces of the no-signaling polytope $\mathrm{NS}(2,2,2)$ of dimension $d=0$ (non-local vertices) and $d=1$ from the Quantum correlation set $\mathrm{Q}(2,2,2)$. Here we continue the proof for other cases. We first show how to exclude $\mathrm{P}$ on the three-dimensional faces of $\mathrm{NS}(2,2,2)$ which automatically includes the faces of lower dimensions and then discuss the $4$-dimensional faces. The PR box we mention in this proof is the one that $p(a,b|x,y) = 1/2$ for $a \oplus b = x \cdot y$.
\begin{enumerate}
    \item [I]
If $\mathrm{P}$ is on the three-dimensional faces of $\mathrm{NS}(2,2,2)$, it can be decomposed as the convex combinations of a PR box and three neighboring Local deterministic boxes:
\begin{equation}\label{face3}
    \mathrm{P}=c_{NS}\cdot PR+ c_{1}\cdot L_i+\cdot c_{2}\cdot L_j + c_3\cdot L_k
\end{equation}
where $i,j,k\in \{1,2\ldots,8\},i\neq j\neq k$ and $c_{NS},c_{1},c_{2},c_3$ are positive and $c_{NS}+c_{1}+c_{2}+c_3=1$. 
Here we need to discuss the following two cases of the local boxes $L_i, L_j, L_k$ in Eq.~\eqref{face3}.\\
(1) Each of $L_i$, $L_j$, $L_k$ contains a non-zero probability event that is also the zero probability event of the PR box, and these events are in the settings $(x_i,y_i),(x_j,y_j),(x_k,y_k)$ that $x_i=x_j=x_k, y_i=y_j\neq y_k$ or $x_i=x_j\neq x_k, y_i=y_j=y_k$. Without loss of generality, we assume $L_i$ is the local box that $p(0|x)=1,p(0|y)=1,\forall x,y\in\{0,1\}$, $L_j$ is the local box that $p(1|x)=1,p(1|y)=1,\forall x,y\in\{0,1\}$ and $L_k$ is the local box that $p(1|x=0)=1,p(0|x=1)=1,p(1|y)=1,\forall y\in\{0,1\}$.

In this case, the matrix $\widetilde{M}_{3,1}$ is constructed as:
	\begin{equation}\label{A21bchain}
        \begin{split}
		&\widetilde{M}_{3,1}=8 \left(|\iota_{(1,2)}\rangle\langle \iota_{(1,2)}|+|\iota_{(2,3)}\rangle\langle \iota_{(2,3)}| +|\iota_{(3,4)}\rangle\langle \iota_{(3,4)}|\right)\\
  &+c_{NS}^2\left(|\iota_{(1)}\rangle\langle\iota_{(1)}|+|\iota_{(4)}\rangle\langle\iota_{(4)}|+2|\iota_{(5)}\rangle\langle\iota_{(5)}|\right)\\
  &+2 c_{NS}\left(|\iota_{(1)}\rangle\langle\iota_{(5)}|+|\iota_{(5)}\rangle\langle\iota_{(1)}|+|\iota_{(4)}\rangle\langle\iota_{(5)}|+|\iota_{(5)}\rangle\langle\iota_{(4)}|\right)\\
        &=\left(
		\begin{array}{ccccc}
			8+c_{NS}^2 & 8 & 0 & 0 & 2c_{NS} \\ 
			8 & 16 & 8 & 0 & 0 \\ 
			0 & 8 & 16 & 8 & 0 \\ 
			0 & 0 & 8 & 8+c_{NS}^2 & 2c_{NS} \\ 
			2c_{NS} & 0 & 0 & 2c_{NS} & 2c_{NS}^2
		\end{array}\right)
        \end{split}
	\end{equation}
	$\widetilde{M}_{3,1}$ can be verified to be positive definite for all $0<c_{NS}<1$ by applying elementary row operations to transform the matrix to an upper triangular matrix with all the diagonal entries positive. We find the five pivots of $\widetilde{M}_{3,1}$ are $c_{NS}^2+8,16-64/(c_{NS}^2+8),(12c_{NS}^2 + 32)/(c_{NS}^2 + 4),(3c_{NS}^4 + 16c_{NS}^2)/(3c_{NS}^2 + 8),(6c_{NS}^4 + 8c_{NS}^2)/(3c_{NS}^2 + 16).$
 
    As before, we relabel the entries of $|\mathrm{P}\rangle$ which is the vector form of the box $\{p(a,b|x,y)\}$, such that the first five entries of $|\mathrm{P}\rangle$ correspond to the probability of events $(1,1|0,1),$ $(0,0|0,0),$ $(1,1|0,0),(0,0|1,0),(1,0|1,1)$. And we see that
	\begin{equation}\label{Aface21}
		\begin{split}
		&\langle \mathrm{P}|\widetilde{M}_{3,1} |\mathrm{P}\rangle-\sum_{i=1}^5 \widetilde{M}_{3,1(i,i)}|\mathrm{P}\>_i\\
        &=c_{NS}^2(|\mathrm{P}\rangle_1^2-|\mathrm{P}\rangle_1+|\mathrm{P}\rangle_4^2-|\mathrm{P}\rangle_4+2|\mathrm{P}\rangle_5^2-2|\mathrm{P}\rangle_5)+4c_{NS}(|\mathrm{P}\rangle_1|\mathrm{P}\rangle_5+|\mathrm{P}\rangle_4|\mathrm{P}\rangle_5)\\
&=c_{NS}^2\left(|\mathrm{P}\rangle_1^2+|\mathrm{P}\rangle_4^2+2|\mathrm{P}\rangle_5^2\right)+c_{NS}^2\left(|\mathrm{P}\rangle_1(\frac{2}{c_{NS}}|\mathrm{P}\rangle_5-1)+|\mathrm{P}\rangle_5(\frac{2}{c_{NS}}|\mathrm{P}\rangle_1-1) \right.\\
&\left. +|\mathrm{P}\rangle_4(\frac{2}{c_{NS}}|\mathrm{P}\rangle_5-1)+|\mathrm{P}\rangle_5(\frac{2}{c_{NS}}|\mathrm{P}\rangle_4-1)\right).
		\end{split}
	\end{equation} 
	Since $|\mathrm{P}\rangle_i> \frac{c_{NS}}{2},\forall i\in \{1,\ldots, 5\}$, we see that \eqref{Aface21} is positive.

    (2) Each of $L_i$, $L_j$, $L_k$ contains a non-zero probability event that is also the zero probability event of the PR box, and these events are in the settings $(x_i,y_i),(x_j,y_j),(x_k,y_k)$ that $x_i=x_j\neq x_k, y_i\neq y_j=y_k$. Without loss of generality, we assume $L_i$ is the local box that $p(0|x)=1,p(0|y)=1,\forall x,y\in\{0,1\}$, $L_j$ is the local box that $p(1|x=0)=1,p(0|x=1)=1,p(1|y)=1,\forall x,y\in\{0,1\}$ and $L_k$ is the local box that $p(0|x=0)=1,p(1|x=1)=1,p(1|y=0)=1,p(0|y=1)=1,\forall y\in\{0,1\}$.
    In this case, the matrix $\widetilde{M}_{3,2}$ is constructed as:
	\begin{equation}
		\begin{split}
			\widetilde{M}_{3,2}&=8\cdot \left(|\iota_{(1,2)}\rangle\langle \iota_{(1,2)}|+|\iota_{(2,3)}\rangle\langle \iota_{(2,3)}|\right)+c_{NS}\left(|\iota_{(1)}\rangle\langle \iota_{(1)}|+|\iota_{(3)}\rangle\langle \iota_{(3)}|\right)\\
   &+(2c_{NS}+c_{NS}^2)\left(|\iota_{(4)}\rangle\langle \iota_{(4)}|+|\iota_{(5)}\rangle\langle \iota_{(5)}|\right)\\
            &+2c_{NS}\left(|\iota_{(1)}\rangle\langle \iota_{(5)}|+|\iota_{(5)}\rangle\langle \iota_{(1)}|+|\iota_{(3)}\rangle\langle \iota_{(4)}|+|\iota_{(4)}\rangle\langle \iota_{(3)}|+|\iota_{(4)}\rangle\langle \iota_{(5)}|+|\iota_{(5)}\rangle\langle \iota_{(4)}|\right)\\
			&=\left(
			\begin{array}{ccccc}
				8+c_{NS} & 8 & 0 & 0 & 2c_{NS} \\ 
				8 & 16 & 8 & 0 & 0 \\ 
				0 & 8 & 8+c_{NS} & 2c_{NS} & 0 \\ 
				0 & 0 & 2c_{NS} & 2c_{NS}+c_{NS}^2 & 2c_{NS} \\ 
				2c_{NS} & 0 & 0 & 2c_{NS} & 2c_{NS}+c_{NS}^2
			\end{array}\right)
		\end{split}
	\end{equation}
	$\widetilde{M}_{3,2}$ can be verified to be positive definite for all $0<c_{NS}<1$ by applying elementary row operations to transform the matrix to an upper triangular matrix with all the diagonal entries positive.  We find the five pivots of $\widetilde{M}_{3,2}$ are $c_{NS}+8,16-64/(c_{NS}+8),(c_{NS}^2 + 8c_{NS})/(c_{NS} + 4),(c_{NS}^3 + 6c_{NS}^2)/(c_{NS} + 8),(c_{NS}^3 + 4c_{NS}^2)/(c_{NS} + 6).$

    Here we relabel the entries of $|\mathrm{P}\rangle$ which is the vector form of the box $\{p(a,b|x,y)\}$, such that the first five entries of $|\mathrm{P}\rangle$ correspond to the probability of events $(1,1|0,1),$ $(0,0|0,1),(1,1|0,0),$ $(0,0|1,0),(1,0|1,1)$, we have
	\begin{equation}\label{Aface3}
		\begin{split}
		  &\langle \mathrm{P}|\widetilde{M}_{3,2} |\mathrm{P}\rangle-\sum_{i=1}^5 \widetilde{M}_{3,2(i,i)}|\mathrm{P}\>_i=c_{NS}\left(|\mathrm{P}\rangle_1^2-|\mathrm{P}\rangle_1+|\mathrm{P}\rangle_3^2-|\mathrm{P}\rangle_3+2|\mathrm{P}\rangle_4^2-2|\mathrm{P}\rangle_4 \right. \\
        & \left. +2|\mathrm{P}\rangle_5^2-2|\mathrm{P}\rangle_5\right)+c_{NS}^2\left(|\mathrm{P}\rangle_4^2-|\mathrm{P}\rangle_4+|\mathrm{P}\rangle_5^2-|\mathrm{P}\rangle_5\right) +4c_{NS}\left(|\mathrm{P}\rangle_1|\mathrm{P}\rangle_5+|\mathrm{P}\rangle_3|\mathrm{P}\rangle_4 \right.\\
        &\left. +|\mathrm{P}\rangle_4|\mathrm{P}\rangle_5\right) =2c_{NS}(|\mathrm{P}\rangle_1+|\mathrm{P}\rangle_4+|\mathrm{P}\rangle_5)(|\mathrm{P}\rangle_1+|\mathrm{P}\rangle_4+|\mathrm{P}\rangle_5-1)\\
        &+c_{NS}^2(|\mathrm{P}\rangle_4^2+|\mathrm{P}\rangle_5^2)-c_{NS}^2(|\mathrm{P}\rangle_4+|\mathrm{P}\rangle_5).
		\end{split}
	\end{equation} 
	Note that $|\mathrm{P}\rangle_1=|\mathrm{P}\rangle_3=c_{NS}/2+c_2$ and $|\mathrm{P}\rangle_1+|\mathrm{P}\rangle_4+|\mathrm{P}\rangle_5=\frac{3}{2}c_{NS}+c_{1}+c_{2}+c_3=1+\frac{1}{2}c_{NS}$, so that \eqref{Aface3} is positive.   

  Thus all the boxes on the 3-dimensional faces of the no-signaling polytope $\mathrm{NS}(2,2,2)$ are excluded from the quantum set $\mathrm{Q}(2,2,2)$ in the same manner. 
  \item [II].
  Now we discuss the $4$-dimensional faces, point $\mathrm{P}$ that are on these faces can be 
    decomposed as the convex combination of one PR box and four neighboring Local deterministic boxes:
    \begin{equation}\label{face4}
    \mathrm{P}=c_{NS}\cdot PR+ c_{1}\cdot L_i+\cdot c_{2}\cdot L_j + c_3\cdot L_k +c_3\cdot L_h
\end{equation}
where $i,j,k,h\in \{1,2\ldots,8\},i\neq j\neq k\neq h$ and $c_{NS},c_{1},c_{2},c_3, c_4$ are positive and $c_{NS}+c_{1}+c_{2}+c_3+c_4=1$. 
Similarly, different relationships of the local boxes $L_i, L_j, L_k, L_h$ in Eq.~\eqref{face4} corresponds to different type of faces of $\mathrm{NS}(2,2,2)$.\\
(1) Each of $L_i$, $L_j$, $L_k, L_h$ contains a non-zero probability event that is also the zero probability event of the PR box, and these events are in the settings $(x_i,y_i),(x_j,y_j),$ $(x_k,y_k),(x_h,y_h)$ that $x_i=x_j,x_k=x_h, y_i=y_j, y_k=y_h$. 
Without loss of generality, we assume $L_i$ is the local box that $p(0|x)=1,p(0|y)=1,\forall x,y\in\{0,1\}$, $L_j$ is the local box that $p(1|x)=1,p(1|y)=1,\forall x,y\in\{0,1\}$, $L_k$ is the local box that $p(0|x=0)=1,p(1|x=1)=1,p(0|y)=1,\forall y\in\{0,1\}$ and $L_h$ is the local box that $p(1|x=0)=1,p(0|x=1)=1,p(1|y)=1,\forall y\in\{0,1\}$. In this case, the matrix $\widetilde{M}_{3,2}$ can be directly used to exclude points $\mathrm{P}$ on this face with the first five entries of $\mathrm{P}$ being the probability of events $(1,1|0,1),(0,0|0,0),(1,1|0,0),(0,0|1,0),(1,0|1,1)$. We have 
    \begin{equation}\label{Aface41}
		\begin{split}
		&	\langle \mathrm{P}|\widetilde{M}_{3,2} |\mathrm{P}\rangle-\sum_{i=1}^5 \widetilde{M}_{3,2(i,i)}|\mathrm{P}\>_i=c_{NS}(|\mathrm{P}\rangle_1^2-|\mathrm{P}\rangle_1+|\mathrm{P}\rangle_3^2-|\mathrm{P}\rangle_3+2|\mathrm{P}\rangle_4^2-2|\mathrm{P}\rangle_4+2|\mathrm{P}\rangle_5^2\\
  &-2|\mathrm{P}\rangle_5)+c_{NS}^2(|\mathrm{P}\rangle_4^2-|\mathrm{P}\rangle_4+|\mathrm{P}\rangle_5^2-|\mathrm{P}\rangle_5)+4c_{NS}(|\mathrm{P}\rangle_1|\mathrm{P}\rangle_5+|\mathrm{P}\rangle_3|\mathrm{P}\rangle_4+|\mathrm{P}\rangle_4|\mathrm{P}\rangle_5)\\
			&=2c_{NS}(|\mathrm{P}\rangle_1+|\mathrm{P}\rangle_4+|\mathrm{P}\rangle_5)(|\mathrm{P}\rangle_1+|\mathrm{P}\rangle_4+|\mathrm{P}\rangle_5-1)\\
   &+c_{NS}^2(|\mathrm{P}\rangle_4^2+|\mathrm{P}\rangle_5^2)-c_{NS}^2(|\mathrm{P}\rangle_4+|\mathrm{P}\rangle_5).
		\end{split}
	\end{equation} 
where $|\mathrm{P}\rangle_1=|\mathrm{P}\rangle_3=c_{NS}/2+c_2+c_4$ and $|\mathrm{P}\rangle_1+|\mathrm{P}\rangle_4+|\mathrm{P}\rangle_5=\frac{3}{2}c_{NS}+c_{1}+c_{2}+c_3+c_4=1+\frac{1}{2}c_{NS}$, so that \eqref{Aface41} is positive. 

(2) Each of $L_i$, $L_j$, $L_k, L_h$ contains a non-zero probability event that is also the zero probability event of the PR box, and these events are in the settings $(x_i,y_i),(x_j,y_j),$ $(x_k,y_k),(x_h,y_h)$ that $x_i=x_j\neq x_k, x_k=x_h, y_i=y_j=y_k\neq y_h$ or $x_i=x_j=x_k\neq x_h, y_i=y_j\neq y_k, y_k=y_h$. 
Similar to case (1), the matrix $\widetilde{M}_{3,2}$ can be directly used to exclude points $\mathrm{P}$ on this face with a suitable labeling of $\mathrm{P}$, we omit the proof here.

(3) Each of $L_i$, $L_j$, $L_k, L_h$ contains a non-zero probability event that is also the zero probability event of the PR box, and these events are in the settings $(x_i,y_i),(x_j,y_j),$ $(x_k,y_k),(x_h,y_h)$ that $x_i=x_j=x_k\neq x_h, y_i=y_j=y_h\neq y_k$. 
Again similar to case (1), the matrix $\widetilde{M}_{3,2}$ can be directly used to exclude points $\mathrm{P}$ on this face with a suitable labeling of $\mathrm{P}$, we omit the proof here.

(4) Each of $L_i$, $L_j$, $L_k, L_h$ contains a non-zero probability event that is also the zero probability event of the PR box, and these events are in the settings $(x_i=0,y_i=0),(x_j=0,y_j=1),(x_k=1,y_k=0),(x_h=1,y_h=1)$. Without loss of generality, we assume $L_i$ is the local box that $p(0|x)=1,p(0|y)=1,\forall x,y\in\{0,1\}$, $L_j$ is the local box that $p(1|x=0)=1,p(0|x=1)=1,p(1|y)=1,\forall x,y\in\{0,1\}$, $L_k$ is the local box that $p(0|x)=1,p(0|y=0)=1,p(1|y=1)=1,\forall x\in\{0,1\}$ and $L_h$ is the local box that $p(0|x=0)=1,p(1|x=1)=1,p(1|y=0)=1,p(0|y=1)=1.$
In this case, we construct a matrix $\widetilde{M}_{4,1}$ as an extension of $\widetilde{M}_{3,2}$ that:
\begin{equation}
    \widetilde{M}_{4,1}=\left(
\begin{array}{ccccc}
8+c_{NS} & 8_{1\times 2} & 0 & 0 & 2c_{NS} \\ 
8_{2\times 1} & 16_{2\times 2} & 8_{2\times 1} & 0_{2\times 1} & 0_{2\times 1} \\ 
0 & 8_{1\times 2} & 8+c_{NS} & 2c_{NS} & 0 \\ 
0 & 0_{1\times 2} & 2c_{NS} & 2c_{NS}+c_{NS}^2 & 2c_{NS} \\ 
2c_{NS} & 0_{1\times 2} & 0 & 2c_{NS} & 2c_{NS}+c_{NS}^2
			\end{array}\right)
\end{equation}
where $8_{1 \times 2}$ refers to a $1 \times 2$ submatrix with all entries being $8$. This matrix $\widetilde{M}_{4,1}$ is positive definite for all $0<c_{NS}<1$ is due to the fact that $\widetilde{M}_{3,2}$ is positive definite. With labeling the first six entries of $\mathrm{P}$ as the probability of events $(1,1|0,0),(0,0|0,0), (0,1|0,0), (1,1|0,1), (1,0|1,1),(0,0|1,0)$ we have
\begin{equation}\label{Aface42}
		\begin{split}
		&	\langle \mathrm{P}|\widetilde{M}_{3,2} |\mathrm{P}\rangle-\sum_{i=1}^5 \widetilde{M}_{3,2(i,i)}|\mathrm{P}\>_i=c_{NS}(|\mathrm{P}\rangle_1^2-|\mathrm{P}\rangle_1+|\mathrm{P}\rangle_4^2-|\mathrm{P}\rangle_4+2|\mathrm{P}\rangle_5^2-2|\mathrm{P}\rangle_5+2|\mathrm{P}\rangle_6^2\\
  &-2|\mathrm{P}\rangle_6)+c_{NS}^2(|\mathrm{P}\rangle_5^2-|\mathrm{P}\rangle_5+|\mathrm{P}\rangle_6^2-|\mathrm{P}\rangle_6)+4c_{NS}(|\mathrm{P}\rangle_1|\mathrm{P}\rangle_6+|\mathrm{P}\rangle_4|\mathrm{P}\rangle_5+|\mathrm{P}\rangle_5|\mathrm{P}\rangle_6)\\
			&=2c_{NS}(|\mathrm{P}\rangle_1+|\mathrm{P}\rangle_5+|\mathrm{P}\rangle_6)(|\mathrm{P}\rangle_1+|\mathrm{P}\rangle_5+|\mathrm{P}\rangle_6-1)\\
   &+c_{NS}^2(|\mathrm{P}\rangle_5^2+|\mathrm{P}\rangle_6^2)-c_{NS}^2(|\mathrm{P}\rangle_5+|\mathrm{P}\rangle_6).
		\end{split}
	\end{equation} 
$|\mathrm{P}\rangle_1=|\mathrm{P}\rangle_4=c_{NS}/2+c_2+c_4$ and $|\mathrm{P}\rangle_1+|\mathrm{P}\rangle_5+|\mathrm{P}\rangle_6=\frac{3}{2}c_{NS}+c_{1}+c_{2}+c_3+c_4=1+\frac{1}{2}c_{NS}$, so that \eqref{Aface42} is positive.

\end{enumerate}
\end{proof}

\section{Aumann's Agreement theorem and its quantum generalization}\label{Asec2}
\subsection*{Proof of theorem~\ref{theo5}}
    Here we prove that these two tables Table~\ref{tab1} and Table~\ref{tab2} are on the $4$-dimensional faces of the $(2,2,2)$ no-signaling polytope. This combined with the theorem~\ref{theo1} in the main text concludes that the Agreement theorems hold for observers of quantum systems.
	
    The proof is quite straightforward, we first rewrite the two tables into our usual form of no-signaling boxes. Then decompose each of these two tables as the convex combination of $1$ PR box and $4$ local deterministic boxes which are neighboring to the PR box. The convex combination coefficients are directly written below each box, and it's clear to see their sum is $1$. The non-negativity of these convex combination coefficients is since the entries in the original tables are non-negative.
	
	Table~\ref{tab1} and Table~\ref{tab2} can be rewritten as the following Tables~\ref{tab1.1} and Table~\ref{tab2.2}:
 
	\vspace{0.5em}
	\hspace{-0.8cm}
		\begin{minipage}{\textwidth}
		\begin{minipage}[t]{0.48\textwidth}
			\makeatletter\def\@captype{table}
			\resizebox{!}{1.1cm}{
				\begin{tabular}{|cc||cc|cc|}
					\hline
					$x\backslash y$&&\multicolumn{2}{c|}{ 0}&\multicolumn{2}{c|}{1}\\
					&$a\backslash b$& \hspace{0.125cm} 0\hspace{0.1cm} &\hspace{0.1cm} 1\hspace{0.125cm}  & \hspace{0.125cm}0 \hspace{0.1cm}& \hspace{0.1cm} 1 \hspace{0.125cm}\\
					\hline \hline
					\multirow{2}*{0} & 0 &    $r$ & $0$ & $r-s$ & $s$  \\
					~ & 1 &        0   & $1-r$ & $-r+t+s$ & $1-t-s$   \\
					\hline 
					\multirow{2}*{0} & 0 &     $t-u$ &  $u$ & $t$ & 0  \\
					~ & 1 &             $r-t+u$ & $1-r-u$ & 0 &  $1-t$  \\\hline
					\end{tabular}
				}
			\caption{Reformatted table of Table~\ref{tab1}. All the entries of the box are non-negative. $r>0$, and $s-u\neq r-t$.}
			\label{tab1.1}
		\end{minipage}
		\quad
		\begin{minipage}[t]{0.48\textwidth}
			\makeatletter\def\@captype{table}
			\resizebox{!}{1.1cm}{
				\begin{tabular}{|cc||cc|cc|}
					\hline
					$x\backslash y$&&\multicolumn{2}{c|}{ 0}&\multicolumn{2}{c|}{1}\\
					&$a\backslash b$& \hspace{0.125cm} 0\hspace{0.1cm} &\hspace{0.1cm} 1\hspace{0.125cm}  & \hspace{0.25cm}0 \hspace{0.1cm}& \hspace{0.1cm} 1 \hspace{0.125cm}\\
					\hline \hline
					\multirow{2}*{0} & 0 &   $s$ & $t$ & 0 & $s+t$    \\
					~ & 1 &                   $1-s-u-t$   & $u$ & $r$ &  $1-s-t-r$ \\
					\hline 
					\multirow{2}*{0} & 0 &    $1-u-t$ & $u+t+r-1$ & $r$ & 0  \\
					~ & 1 &             0 & $1-r$ & 0 & $1-r$    \\\hline
				\end{tabular}
				}
			\caption{Reformatted table of Table~\ref{tab2}. All the entries of the box are non-negative. $s>0$, and $s+t \neq 0$ and $u+t \neq 1$.}
			\label{tab2.2}
		\end{minipage}
	\end{minipage}
	
	\begin{itemize}
		\item [1.] Decomposition of Table~\ref{tab1.1} with $s-u> r-t$. \\  

  \hspace{-0.8cm}
		\begin{minipage}{\textwidth}
			\begin{minipage}[t]{0.32\textwidth}
				\makeatletter\def\@captype{table}
				\resizebox{!}{1.1cm}{
					\begin{tabular}{|cc||cc|cc|}
						\hline
					$x\backslash y$&&\multicolumn{2}{c|}{ 0}&\multicolumn{2}{c|}{1}\\
					&$a\backslash b$& \hspace{0.1cm} 0\hspace{0.1cm} &\hspace{0.1cm} 1\hspace{0.1cm}  & \hspace{0.1cm}0 \hspace{0.1cm}& \hspace{0.1cm} 1 \hspace{0.1cm}\\
					\hline \hline
					\multirow{2}*{0} & 0 &   1/2 & 0 & 0 & 1/2    \\
					~ & 1 &                   0   & 1/2 & 1/2& 0  \\
					\hline 
					\multirow{2}*{0} & 0 &     1/2& 0 & 1/2 &0   \\
					~ & 1 &                    0  &1/2 & 0 & 1/2   \\\hline
					\end{tabular}
				}
				\caption{$2(s+t-r-u)$}
			\end{minipage}
			\begin{minipage}[t]{0.32\textwidth}
				\makeatletter\def\@captype{table}
				\resizebox{!}{1.1cm}{
					\begin{tabular}{|cc||cc|cc|}
						\hline
						$x\backslash y$&&\multicolumn{2}{c|}{ 0}&\multicolumn{2}{c|}{1}\\
						&$a\backslash b$& \hspace{0.1cm} 0\hspace{0.1cm} &\hspace{0.1cm} 1\hspace{0.1cm}  & \hspace{0.1cm}0 \hspace{0.1cm}& \hspace{0.1cm} 1 \hspace{0.1cm}\\
						\hline \hline
						\multirow{2}*{0} & 0 &   1 & 0 & 1 &0    \\
						~ & 1 &                   0   & 0 & 0& 0  \\
						\hline 
						\multirow{2}*{0} & 0 &     1&0  &1  & 0  \\
						~ & 1 &                     0 &0 &0  &   0 \\\hline
					\end{tabular}
				}
				\caption{$r-s$}
			\end{minipage}
			\begin{minipage}[t]{0.32\textwidth}
				\makeatletter\def\@captype{table}
				\resizebox{!}{1.1cm}{
					\begin{tabular}{|cc||cc|cc|}
						\hline
						$x\backslash y$&&\multicolumn{2}{c|}{ 0}&\multicolumn{2}{c|}{1}\\
						&$a\backslash b$& \hspace{0.1cm} 0\hspace{0.1cm} &\hspace{0.1cm} 1\hspace{0.1cm}  & \hspace{0.1cm}0 \hspace{0.1cm}& \hspace{0.1cm} 1 \hspace{0.1cm}\\
						\hline \hline
						\multirow{2}*{0} & 0 &   1 & 0 & 0 & 1   \\
						~ & 1 &                    0  &  0& 0&  0 \\
						\hline 
						\multirow{2}*{0} & 0 &    0 & 0 &  0& 0  \\
						~ & 1 &                     1 & 0&0  &  1  \\\hline
					\end{tabular}
				}
				\caption{$r-t+u$}
			\end{minipage}
			\begin{minipage}[t]{0.32\textwidth}
				\makeatletter\def\@captype{table}
				\resizebox{!}{1.1cm}{
					\begin{tabular}{|cc||cc|cc|}
						\hline
						$x\backslash y$&&\multicolumn{2}{c|}{ 0}&\multicolumn{2}{c|}{1}\\
						&$a\backslash b$& \hspace{0.1cm} 0\hspace{0.1cm} &\hspace{0.1cm} 1\hspace{0.1cm}  & \hspace{0.1cm}0 \hspace{0.1cm}& \hspace{0.1cm} 1 \hspace{0.1cm}\\
						\hline \hline
						\multirow{2}*{0} & 0 &  0  & 0 & 0 & 0   \\
						~ & 1 &                    0  & 1 &1 & 0  \\
						\hline 
						\multirow{2}*{0} & 0 &   0  &  1&  1&  0 \\
						~ & 1 &                    0  &0 & 0 &   0 \\\hline
					\end{tabular}
				}
				\caption{$u$}
			\end{minipage}
			\begin{minipage}[t]{0.32\textwidth}
				\makeatletter\def\@captype{table}
				\resizebox{!}{1.1cm}{
					\begin{tabular}{|cc||cc|cc|}
						\hline
						$x\backslash y$&&\multicolumn{2}{c|}{ 0}&\multicolumn{2}{c|}{1}\\
						&$a\backslash b$& \hspace{0.1cm} 0\hspace{0.1cm} &\hspace{0.1cm} 1\hspace{0.1cm}  & \hspace{0.1cm}0 \hspace{0.1cm}& \hspace{0.1cm} 1 \hspace{0.1cm}\\
						\hline \hline
						\multirow{2}*{0} & 0 &   0 & 0 & 0 & 0   \\
						~ & 1 &                   0   & 1 & 0& 1  \\
						\hline 
						\multirow{2}*{0} & 0 &   0  &0  & 0 &  0 \\
						~ & 1 &                   0   & 1& 0 &1    \\\hline
					\end{tabular}
				}
				\caption{$1-t-s$}
			\end{minipage}
		\end{minipage}
		
		\item [2.] Decomposition of Table~\ref{tab1.1} with $s-u< r-t$.\\

  \hspace{-0.8cm}
		\begin{minipage}{\textwidth}
			\begin{minipage}[t]{0.32\textwidth}
				\makeatletter\def\@captype{table}
				\resizebox{!}{1.1cm}{
					\begin{tabular}{|cc||cc|cc|}
						\hline
						$x\backslash y$&&\multicolumn{2}{c|}{ 0}&\multicolumn{2}{c|}{1}\\
						&$a\backslash b$& \hspace{0.1cm} 0\hspace{0.1cm} &\hspace{0.1cm} 1\hspace{0.1cm}  & \hspace{0.1cm}0 \hspace{0.1cm}& \hspace{0.1cm} 1 \hspace{0.1cm}\\
						\hline \hline
						\multirow{2}*{0} & 0 &  1/2  &0  &1/2  &0    \\
						~ & 1 &                   0   & 1/2 &0 &1/2   \\
						\hline 
						\multirow{2}*{0} & 0 &    0 &1/2  &1/2  &0   \\
						~ & 1 &                   1/2   &0 & 0 &1/2    \\\hline
					\end{tabular}
				}
				\caption{$2(u+r-t-s)$}
			\end{minipage}
			\begin{minipage}[t]{0.32\textwidth}
				\makeatletter\def\@captype{table}
				\resizebox{!}{1.1cm}{
					\begin{tabular}{|cc||cc|cc|}
						\hline
						$x\backslash y$&&\multicolumn{2}{c|}{ 0}&\multicolumn{2}{c|}{1}\\
						&$a\backslash b$& \hspace{0.1cm} 0\hspace{0.1cm} &\hspace{0.1cm} 1\hspace{0.1cm}  & \hspace{0.1cm}0 \hspace{0.1cm}& \hspace{0.1cm} 1 \hspace{0.1cm}\\
						\hline \hline
						\multirow{2}*{0} & 0 &   1 & 0 & 1 &  0  \\
						~ & 1 &                    0  &0  & 0& 0  \\
						\hline 
						\multirow{2}*{0} & 0 &     1& 0 & 1 & 0  \\
						~ & 1 &                     0 & 0& 0 &  0  \\\hline
					\end{tabular}
				}
				\caption{$t-u$}
			\end{minipage}
			\begin{minipage}[t]{0.32\textwidth}
				\makeatletter\def\@captype{table}
				\resizebox{!}{1.1cm}{
					\begin{tabular}{|cc||cc|cc|}
						\hline
						$x\backslash y$&&\multicolumn{2}{c|}{ 0}&\multicolumn{2}{c|}{1}\\
						&$a\backslash b$& \hspace{0.1cm} 0\hspace{0.1cm} &\hspace{0.1cm} 1\hspace{0.1cm}  & \hspace{0.1cm}0 \hspace{0.1cm}& \hspace{0.1cm} 1 \hspace{0.1cm}\\
						\hline \hline
						\multirow{2}*{0} & 0 &   1 & 0 & 0 &  1  \\
						~ & 1 &                   0   & 0 &0 &  0 \\
						\hline 
						\multirow{2}*{0} & 0 &    0 & 0 & 0 & 0  \\
						~ & 1 &                    1  &0 & 0 & 1   \\\hline
					\end{tabular}
				}
				\caption{$s$}
			\end{minipage}
			\begin{minipage}[t]{0.32\textwidth}
				\makeatletter\def\@captype{table}
				\resizebox{!}{1.1cm}{
					\begin{tabular}{|cc||cc|cc|}
						\hline
						$x\backslash y$&&\multicolumn{2}{c|}{ 0}&\multicolumn{2}{c|}{1}\\
						&$a\backslash b$& \hspace{0.1cm} 0\hspace{0.1cm} &\hspace{0.1cm} 1\hspace{0.1cm}  & \hspace{0.1cm}0 \hspace{0.1cm}& \hspace{0.1cm} 1 \hspace{0.1cm}\\
						\hline \hline
						\multirow{2}*{0} & 0 &    0& 0 & 0 & 0   \\
						~ & 1 &                    0  & 1 & 1& 0  \\
						\hline 
						\multirow{2}*{0} & 0 &    0 & 1 & 1 & 0  \\
						~ & 1 &                    0  &0 & 0 &  0  \\\hline
					\end{tabular}
				}
				\caption{$-r+t+s$}
			\end{minipage}
			\begin{minipage}[t]{0.32\textwidth}
				\makeatletter\def\@captype{table}
				\resizebox{!}{1.1cm}{
					\begin{tabular}{|cc||cc|cc|}
						\hline
						$x\backslash y$&&\multicolumn{2}{c|}{ 0}&\multicolumn{2}{c|}{1}\\
						&$a\backslash b$& \hspace{0.1cm} 0\hspace{0.1cm} &\hspace{0.1cm} 1\hspace{0.1cm}  & \hspace{0.1cm}0 \hspace{0.1cm}& \hspace{0.1cm} 1 \hspace{0.1cm}\\
						\hline \hline
						\multirow{2}*{0} & 0 &    0& 0 & 0 & 0   \\
						~ & 1 &                    0  & 1 & 0& 1  \\
						\hline 
						\multirow{2}*{0} & 0 &    0 & 0 & 0 &  0 \\
						~ & 1 &                    0  &1&0  &  1  \\\hline
					\end{tabular}
				}
				\caption{$1-r-u$}
			\end{minipage}
		\end{minipage}
		\item [3.] Decomposition of Table~\ref{tab2.2}.\\
  
  \hspace{-0.8cm}
		\begin{minipage}{\textwidth}
			\begin{minipage}[t]{0.32\textwidth}
				\makeatletter\def\@captype{table}
				\resizebox{!}{1.1cm}{
					\begin{tabular}{|cc||cc|cc|}
						\hline
						$x\backslash y$&&\multicolumn{2}{c|}{ 0}&\multicolumn{2}{c|}{1}\\
						&$a\backslash b$& \hspace{0.1cm} 0\hspace{0.1cm} &\hspace{0.1cm} 1\hspace{0.1cm}  & \hspace{0.1cm}0 \hspace{0.1cm}& \hspace{0.1cm} 1 \hspace{0.1cm}\\
						\hline \hline
						\multirow{2}*{0} & 0 &  1/2  & 0 & 0 & 1/2   \\
						~ & 1 &                   0   & 1/2 & 1/2 & 0  \\
						\hline 
						\multirow{2}*{0} & 0 &   1/2  & 0 & 1/2 & 0  \\
						~ & 1 &                   0   & 1/2 & 0 & 1/2    \\\hline
					\end{tabular}
				}
				\caption{$2s$}
			\end{minipage}
			\begin{minipage}[t]{0.32\textwidth}
				\makeatletter\def\@captype{table}
				\resizebox{!}{1.1cm}{
					\begin{tabular}{|cc||cc|cc|}
						\hline
						$x\backslash y$&&\multicolumn{2}{c|}{ 0}&\multicolumn{2}{c|}{1}\\
						&$a\backslash b$& \hspace{0.1cm} 0\hspace{0.1cm} &\hspace{0.1cm} 1\hspace{0.1cm}  & \hspace{0.1cm}0 \hspace{0.1cm}& \hspace{0.1cm} 1 \hspace{0.1cm}\\
						\hline \hline
						\multirow{2}*{0} & 0 &   0 & 0 & 0 &0    \\
						~ & 1 &                    0  &  1& 1&  0 \\
						\hline 
						\multirow{2}*{0} & 0 &    0 & 1 & 1 & 0  \\
						~ & 1 &                    0  & 0& 0 &  0  \\\hline
					\end{tabular}
				}
				\caption{$u+t+r-1$}
			\end{minipage}
			\begin{minipage}[t]{0.32\textwidth}
				\makeatletter\def\@captype{table}
				\resizebox{!}{1.1cm}{
					\begin{tabular}{|cc||cc|cc|}
						\hline
						$x\backslash y$&&\multicolumn{2}{c|}{ 0}&\multicolumn{2}{c|}{1}\\
						&$a\backslash b$& \hspace{0.1cm} 0\hspace{0.1cm} &\hspace{0.1cm} 1\hspace{0.1cm}  & \hspace{0.1cm}0 \hspace{0.1cm}& \hspace{0.1cm} 1 \hspace{0.1cm}\\
						\hline \hline
						\multirow{2}*{0} & 0 &   0 & 0 &  0& 0   \\
						~ & 1 &                   1   & 0 & 1& 0  \\
						\hline 
						\multirow{2}*{0} & 0 &    1 & 0 & 1 & 0  \\
						~ & 1 &                    0  &0 & 0 &   0 \\\hline
					\end{tabular}
				}
				\caption{$1-s-u-t$}
			\end{minipage}
			\begin{minipage}[t]{0.32\textwidth}
				\makeatletter\def\@captype{table}
				\resizebox{!}{1.1cm}{
					\begin{tabular}{|cc||cc|cc|}
						\hline
						$x\backslash y$&&\multicolumn{2}{c|}{ 0}&\multicolumn{2}{c|}{1}\\
						&$a\backslash b$& \hspace{0.1cm} 0\hspace{0.1cm} &\hspace{0.1cm} 1\hspace{0.1cm}  & \hspace{0.1cm}0 \hspace{0.1cm}& \hspace{0.1cm} 1 \hspace{0.1cm}\\
						\hline \hline
						\multirow{2}*{0} & 0 &  0  &  0& 0 & 0   \\
						~ & 1 &                    0  & 1 & 0 & 1  \\
						\hline 
						\multirow{2}*{0} & 0 &    0 & 0 & 0 & 0  \\
						~ & 1 &                    0  & 1 & 0 & 1   \\\hline
					\end{tabular}
				}
				\caption{$1-s-t-r$}
			\end{minipage}
			\begin{minipage}[t]{0.32\textwidth}
				\makeatletter\def\@captype{table}
				\resizebox{!}{1.1cm}{
					\begin{tabular}{|cc||cc|cc|}
						\hline
						$x\backslash y$&&\multicolumn{2}{c|}{ 0}&\multicolumn{2}{c|}{1}\\
						&$a\backslash b$& \hspace{0.1cm} 0\hspace{0.1cm} &\hspace{0.1cm} 1\hspace{0.1cm}  & \hspace{0.1cm}0 \hspace{0.1cm}& \hspace{0.1cm} 1 \hspace{0.1cm}\\
						\hline \hline
						\multirow{2}*{0} & 0 &  0  & 1 & 0 &1    \\
						~ & 1 &                    0  & 0 & 0& 0  \\
						\hline 
						\multirow{2}*{0} & 0 &    0 & 0 & 0 & 0  \\
						~ & 1 &                    0  &1 & 0 & 1   \\\hline
					\end{tabular}
				}
				\caption{$t$}
			\end{minipage}
		\end{minipage}
		
	\end{itemize}

\section{Common faces of the Quantum and Classical Correlation sets}\label{Asec4}
In this section, we show proofs of lemmas we used in the main text, construction of the game matrix $\G_{2^k}$ for the special case that $m_a = m_b = 2^k$ with $k \geq 2$, and prove that $\G_{2^k}$ is diagonal in the Hadamard basis which reproduces the class of games discovered by Linden et al. in \cite{linden2007quantum}

\begin{lemma}{\label{Alem_unique}}
Let $m_a,m_b\in\mathbb{N}$ and $\mathbb{N}\ni r\leq \log_2(m_a+m_b)$. Let $\{v_1,...,v_r\}$ be set of general positions vector of $2^{m_a+m_b}$ rows. Then up to swapping rows and re-indexing, the first $2^r$ rows are unique. In particular, the unique choice for the first $2^r$ rows is the lexicographical ordering. 

\end{lemma}
\begin{proof}
Denote $(v_j)_i$ to be the $i^{th}$ row of the vector $v_j$. Fix $I\subseteq \{1,...,r\}$. Since $\odot_j v_{I,j}\neq 0$, there exists $i\in\{1,...2^r\}$ such that $\prod_j(v_{I,j})_i\neq 0$. Thus $(v_{I,j})_i\neq 0$ for all $j\in\{1,...,r\}$. Therefore, we have $(v_j)_i=1$ if $j\in I$ and $(v_j)_i=-1$ if $j\not\in I$. Thus for every choice of $I$, there is a unique row setting such that it ensures $\odot_j v_{I,j}\neq 0$. Since we have $r$ vectors and we need $2^r$ difference row settings, up to swapping rows and re-indexing $v_i$'s, the first $2^r$ rows are in a lexicographical ordering. In other words, up to swapping rows and re-indexing $v_i$'s, the first $2^r$ rows are unique and are ordered by lexicographical ordering.
\end{proof}

By Lemma \ref{Alem_unique}, if $m_a=m_b=2^k$, where $k\geq 2$, $r=\log_2(m_a+m_b)=k+1$ and $\{v_1,...,v_r\}$ are in general positions, then $\{v_1,...,v_r\}$ are unique up to swapping rows and re-indexing. We now focus on this case. For convenience reason, we first concatenate $v_1,...,v_r$ into one matrix, denoted by $G_{2^k}$. Notice that $G_{2^k}$ is a $2^{k+1}$ by $r$ matrix. We then define the $2^k$ by $2^k$ game matrix $\G_{2^k}$ as follows. The $(i,j)$ position of $\G_{2^k}$ associates with Alice's $i^{th}$ input and Bob's $j^{th}$ input. In other words, the $(i,j)$ position of $\G_{2^k}$ associates with the $i^{th}$ row and the ${(2^k+j)}^{th}$ row of $G_{2^k}$ respectively. Let $r_i=(x_1,...,x_r)$ and $r_j=(y_1,...,y_r)$ be the $i^{th}$ row and the $(2^k+j)^{th}$ row of $G_{2^k}$ respectively. We define the $(i,j)$ entry of $\G_{2^k}$ be using an operator $\star$ on the $i^{th}$ row and the $(2^k+j)^{th}$ row of $G_{2^k}$ as follows:
$$
(\G_{2^k})_{i,j}=r_i\star r_j=\begin{cases}
1 &\text{if $x_1y_1+...+x_ry_r>0$}\\
-1 &\text{if $x_1y_1+...+x_ry_r<0$}\\
0 &\text{if $x_1y_1+...+x_ry_r=0$}\\
\end{cases}
$$

\begin{examp}{\label{Aex}}
If $k=2$ (thus $m_a=m_b=2^2=4$ and $r=3$), we have
$$v_1=\begin{pmatrix}1\\1\\1\\1\\-1\\-1\\-1\\-1\end{pmatrix};
v_2 = \begin{pmatrix}1\\1\\-1\\-1\\1\\1\\-1\\-1\end{pmatrix};
v_3 = \begin{pmatrix}1\\-1\\1\\-1\\1\\-1\\1\\-1\end{pmatrix}; 
G_{2^2}=\begin{pmatrix}1&1&1\\1&1&-1\\1&-1&1\\1&-1&-1\\-1&1&1\\-1&1&-1\\-1&-1&1\\-1&-1&-1\end{pmatrix};
\G_{2^2}=\begin{pmatrix}1&-1&-1&-1\\-1&1&-1&-1\\-1&-1&1&-1\\-1&-1&-1&1\end{pmatrix};
$$
for example, we have $(\G_{2^2})_{1,1}=(1,1,1)\star (-1,1,1)=1$ and  $(\G_{2^2})_{1,2}=(1,1,1)\star (-1,1,-1)=-1$.
\end{examp}
If $k=3$ (thus $m_a=m_b=2^3=8$ and $r=4$), we have
$$v_1=\begin{pmatrix}1\\1\\1\\1\\1\\1\\1\\1\\-1\\-1\\-1\\-1\\-1\\-1\\-1\\-1\end{pmatrix};
v_2=\begin{pmatrix}1\\1\\1\\1\\-1\\-1\\-1\\-1\\1\\1\\1\\1\\-1\\-1\\-1\\-1\end{pmatrix};
v_3 = \begin{pmatrix}1\\1\\-1\\-1\\1\\1\\-1\\-1\\1\\1\\-1\\-1\\1\\1\\-1\\-1\end{pmatrix};
v_4 = \begin{pmatrix}1\\-1\\1\\-1\\1\\-1\\1\\-1\\1\\-1\\1\\-1\\1\\-1\\1\\-1\end{pmatrix}; 
G_{2^3}=\begin{pmatrix}1&1&1&1\\
1&1&1&-1\\
1&1&-1&1\\
1&1&-1&-1\\
1&-1&1&1\\
1&-1&1&-1\\
1&-1&-1&1\\
1&-1&-1&-1\\
-1&1&1&1\\
-1&1&1&-1\\
-1&1&-1&1\\
-1&1&-1&-1\\
-1&-1&1&1\\
-1&-1&1&-1\\
-1&-1&-1&1\\
-1&-1&-1&-1\end{pmatrix};
$$
and
$$\G_{2^3}=\begin{pmatrix}
1&0&0&-1&0&-1&-1&-1\\
0&1&-1&0&-1&0&-1&-1\\
0&-1&1&0&-1&-1&0&-1\\
-1&0&0&1&-1&-1&-1&0\\
0&-1&-1&-1&1&0&0&-1\\
-1&0&-1&-1&0&1&-1&0\\
-1&-1&0&-1&0&-1&1&0\\
-1&-1&-1&0&-1&0&0&1
\end{pmatrix}.$$

\begin{definition}
Define $g_k:\{1,...,2^k\}\to\{0,1\}^k$ by sending $i\in\{1,...,2^k\}$ to binary representation of $i-1$. Define $f_k:\{0,1\}^k\to\{\pm1\}^k$ by sending $0$ to $1$ and $1$ to $-1$. Let $i\in\{1,...,2^k\}$, we denote $\tilde{i_k}=f_k\circ g_k(i)$.

For example, $g_3(2)=(0,0,1)$, $f_3(0,0,1)=(1,1,-1)$ and $\tilde{2_3}=(1,1,-1)$.
\end{definition}
By above definition, we can observe that $(\G_{2^k})_{i,j}=(1,\tilde{i}_k)\star(-1,\tilde{j}_k)$.
\begin{definition}
Let $k,x$ be positive integers. Define $\G_{2^k,x}$ to be the top left corner square block of $\G_{2^k}$ with dimension $x$ by $x$. Define $\overline{\G_{2^k,x}}$ is the top-most right-most corner square block of $\G_{2^k}$ of dimension $x$ by $x$. In particular, $\G_{2^k,2^k}=\overline{\G_{2^k,2^k}}=\G_{2^k}$.

For example, $\G_{2^2,1}=(1)$, $\overline{\G_{2^2,1}}=(-1)$, $\G_{2^2,2}=\begin{pmatrix}1&-1\\-1&1\end{pmatrix}$ and $\overline{\G_{2^2,2}}=\begin{pmatrix}-1&-1\\-1&-1\end{pmatrix}$
\end{definition}

\begin{lemma}{\label{Alem_main}}
Consider the game matrix $\G_{2^k}$, where $k\geq 4$ and denote $d=dim(\G_{2^k})=2^k$. The game matrix $\G_{2^k}$ has the form
$$
\G_{2^k}=\begin{pmatrix}
	A^{(d)}_{2^{k-2}}&\G_{2^{k-2}}&\G_{2^{k-2}}&B^{(d)}_{2^{k-2}}\\
	\G_{2^{k-2}}&A^{(d)}_{2^{k-2}}&B^{(d)}_{2^{k-2}}&\G_{2^{k-2}}\\
	\G_{2^{k-2}}&B^{(d)}_{2^{k-2}}&A^{(d)}_{2^{k-2}}&\G_{2^{k-2}}\\
	B^{(d)}_{2^{k-2}}&\G_{2^{k-2}}&\G_{2^{k-2}}&A^{(d)}_{2^{k-2}}
\end{pmatrix}
$$
where $A^{(d)}_{2^{k-2}}=\begin{pmatrix}A^{(d)}_{2^{k-3}}&\G_{\frac{d}{4},2^{k-3}}\\\G_{\frac{d}{4},2^{k-3}}&A^{(d)}_{2^{k-3}}\end{pmatrix}$, $B^{(d)}_{2^{k-2}}=\begin{pmatrix}\overline{\G_{\frac{d}{4},2^{k-3}}}&B^{(d)}_{2^{k-3}}\\B^{(d)}_{2^{k-3}}&\overline{\G_{\frac{d}{4},2^{k-3}}}\end{pmatrix}$, $A^{(d)}_1=1$ and $B^{(d)}_1=-1$.
\end{lemma}
\begin{proof}
We first partition $\G_{2^k}$ into 16 sub square block matrices and label them as 
$$\G_{2^k}=
\begin{pmatrix}
	M_{1,1}&M_{1,2}&M_{1,3}&M_{1,4}\\
	M_{2,1}&M_{2,2}&M_{2,3}&M_{2,4}\\
	M_{3,1}&M_{3,2}&M_{3,3}&M_{3,4}\\
	M_{4,1}&M_{4,2}&M_{4,3}&M_{4,4}\\
\end{pmatrix}$$
First, we want to show that $M_{1,2}=M_{1,3}=M_{2,1}=M_{2,4}=M_{3,1}=M_{3,4}=M_{4,2}=M_{4,3}=\G_{2^{k-2}}$. For $i,j\in\{1,...,2^{k-2}\}$, we have
\begin{align*}
	(M_{1,2})_{i,j}&=(1,1,1,\tilde{i}_{k-2})\star(-1,1,-1,\tilde{j}_{k-2})=(1,\cancel{1,1},\tilde{i}_{k-2})\star(-1,\cancel{1,-1},\tilde{j}_{k-2})\\
 &=(1,\tilde{i}_{k-2})\star(-1,\tilde{j}_{k-2})=(\G_{2^{k-2}})_{i,j}\\
	(M_{1,3})_{i,j}&=(1,1,1,\tilde{i}_{k-2})\star(-1,-1,1,\tilde{j}_{k-2})=(1,\cancel{1,1},\tilde{i}_{k-2})\star(-1,\cancel{-1,1},\tilde{j}_{k-2})\\
 &=(1,\tilde{i}_{k-2})\star(-1,\tilde{j}_{k-2})=(\G_{2^{k-2}})_{i,j}\\
	(M_{2,1})_{i,j}&=(1,1,-1,\tilde{i}_{k-2})\star(-1,1,1,\tilde{j}_{k-2})=(1,\cancel{1,-1},\tilde{i}_{k-2})\star(-1,\cancel{1,1},\tilde{j}_{k-2})\\
 &=(1,\tilde{i}_{k-2})\star(-1,\tilde{j}_{k-2})=(\G_{2^{k-2}})_{i,j}\\
	(M_{2,4})_{i,j}&=(1,1,-1,\tilde{i}_{k-2})\star(-1,-1,-1,\tilde{j}_{k-2})=(1,\cancel{1,-1},\tilde{i}_{k-2})\star(-1,\cancel{-1,-1},\tilde{j}_{k-2})\\
 &=(1,\tilde{i}_{k-2})\star(-1,\tilde{j}_{k-2})=(\G_{2^{k-2}})_{i,j}\\
	(M_{3,1})_{i,j}&=(1,-1,1,\tilde{i}_{k-2})\star(-1,1,1,\tilde{j}_{k-2})=(1,-\cancel{1,1},\tilde{i}_{k-2})\star(-1,\cancel{1,1},\tilde{j}_{k-2})\\
 &=(1,\tilde{i}_{k-2})\star(-1,\tilde{j}_{k-2})=(\G_{2^{k-2}})_{i,j}\\
	(M_{3,4})_{i,j}&=(1,-1,1,\tilde{i}_{k-2})\star(-1,-1,-1,\tilde{j}_{k-2})=(1,\cancel{-1,1},\tilde{i}_{k-2})\star(-1,\cancel{-1,-1},\tilde{j}_{k-2})\\
 &=(1,\tilde{i}_{k-2})\star(-1,\tilde{j}_{k-2})=(\G_{2^{k-2}})_{i,j}\\
	(M_{4,2})_{i,j}&=(1,-1,-1,\tilde{i}_{k-2})\star(-1,-1,1,\tilde{j}_{k-2})=(1,\cancel{-1,-1},\tilde{i}_{k-2})\star(-1,\cancel{-1,1},\tilde{j}_{k-2})\\
 &=(1,\tilde{i}_{k-2})\star(-1,\tilde{j}_{k-2})=(\G_{2^{k-2}})_{i,j}\\
	(M_{4,3})_{i,j}&=(1,-1,-1,\tilde{i}_{k-2})\star(-1,-1,1,\tilde{j}_{k-2})=(1,\cancel{-1,-1},\tilde{i}_{k-2})\star(-1,\cancel{-1,1},\tilde{j}_{k-2})\\
 &=(1,\tilde{i}_{k-2})\star(-1,\tilde{j}_{k-2})=(\G_{2^{k-2}})_{i,j}\\
\end{align*}
Therefore $M_{1,2}=M_{1,3}=M_{2,1}=M_{2,4}=M_{3,1}=M_{3,4}=M_{4,2}=M_{4,3}=\G_{2^{k-2}}$. Next, we consider $M_{1,1},M_{2,2},M_{3,3}$ and $M_{4,4}$. For $i,j\in\{1,...,2^{k-2}\}$, we have
\begin{align*}
	(M_{1,1})_{i,j}&=(1,1,1,\tilde{i}_{k-2})\star(-1,1,1,\tilde{j}_{k-2})\\
	(M_{2,2})_{i,j}&=(1,1,-1,\tilde{i}_{k-2})\star(-1,1,-1,\tilde{j}_{k-2})=(1,1,1,\tilde{i}_{k-2})\star(-1,1,1,\tilde{j}_{k-2})=(M_{1,1})_{i,j}\\
	(M_{3,3})_{i,j}&=(1,-1,1,\tilde{i}_{k-2})\star(-1,-1,1,\tilde{j}_{k-2})=(1,1,1,\tilde{i}_{k-2})\star(-1,1,1,\tilde{j}_{k-2})=(M_{1,1})_{i,j}\\
	(M_{4,4})_{i,j}&=(1,-1,-1,\tilde{i}_{k-2})\star(-1,-1,-1,\tilde{j}_{k-2})=(1,1,1,\tilde{i}_{k-2})\star(-1,1,1,\tilde{j}_{k-2})=(M_{1,1})_{i,j}
\end{align*}
Therefore $M_{1,1}=M_{2,2}=M_{3,3}=M_{4,4}$. After that, consider $M_{1,4},M_{2,3},M_{3,2}$ and $M_{4,1}$. For $i,j\in\{1,...,2^{k-2}\}$, we have
\begin{align*}
	(M_{1,4})_{i,j}&=(1,1,1,\tilde{i}_{k-2})\star(-1,-1,-1,\tilde{j}_{k-2})\\
	(M_{2,3})_{i,j}&=(1,1,-1,\tilde{i}_{k-2})\star(-1,-1,1,\tilde{j}_{k-2})=(1,1,1,\tilde{i}_{k-2})\star(-1,-1,-1,\tilde{j}_{k-2})=(M_{1,4})_{i,j}\\
	(M_{3,2})_{i,j}&=(1,-1,1,\tilde{i}_{k-2})\star(-1,1,-1,\tilde{j}_{k-2})=(1,1,1,\tilde{i}_{k-2})\star(-1,-1,-1,\tilde{j}_{k-2})=(M_{1,4})_{i,j}\\
	(M_{4,1})_{i,j}&=(1,-1,-1,\tilde{i}_{k-2})\star(-1,1,1,\tilde{j}_{k-2})=(1,1,1,\tilde{i}_{k-2})\star(-1,-1,-1,\tilde{j}_{k-2})=(M_{1,4})_{i,j}
\end{align*}
Thus $M_{1,4}=M_{2,3}=M_{3,2}=M_{4,1}$. Hence we can conclude that
$$
\G_{2^k}=\begin{pmatrix}
	A^{(d)}_{2^{k-2}}&\G_{2^{k-2}}&\G_{2^{k-2}}&B^{(d)}_{2^{k-2}}\\
	\G_{2^{k-2}}&A^{(d)}_{2^{k-2}}&B^{(d)}_{2^{k-2}}&\G_{2^{k-2}}\\
	\G_{2^{k-2}}&B^{(d)}_{2^{k-2}}&A^{(d)}_{2^{k-2}}&\G_{2^{k-2}}\\
	B^{(d)}_{2^{k-2}}&\G_{2^{k-2}}&\G_{2^{k-2}}&A^{(d)}_{2^{k-2}}
\end{pmatrix}
$$
for some $2^{k-2}\times 2^{k-2}$ matrices $A^{(d)}_{2^{k-2}}$ and $B^{(d)}_{2^{k-2}}$. We remain to consider matrix $A^{(d)}_{2^{k-2}}$ and $B^{(d)}_{2^{k-2}}$. 

First, we consider $A^{(d)}_{2^{k-2}}$ and partition it into four sub square block matrices as $A^{(d)}_{2^{k-2}}=\begin{pmatrix}
	N_{1,1}&N_{1,2}\\N_{2,1}&N_{2,2}
\end{pmatrix}$. For $i,j\in\{1,...,2^{k-3}\}$, we have
\begin{align*}
	(N_{1,2})_{i,j}&=(1,1,1,1,\tilde{i}_{k-3})\star(-1,1,1,-1,\tilde{j}_{k-3})\\
 &=(1,1,1,-1,\tilde{i}_{k-3})\star(-1,1,1,1,\tilde{j}_{k-3})=(N_{2,1})_{i,j}\\
	(N_{1,1})_{i,j}&=(1,1,1,1,\tilde{i}_{k-3})\star(-1,1,1,1,\tilde{j}_{k-3})\\
 &=(1,1,1,-1,\tilde{i}_{k-3})\star(-1,1,1,-1,\tilde{j}_{k-3})=(N_{2,2})_{i,j}\\
	(N_{1,2})_{i,j}&=(1,1,1,1,\tilde{i}_{k-3})\star(-1,1,1,-1,\tilde{j}_{k-3})=(1,1,\cancel{1,1},\tilde{i}_{k-3})\star(-1,1,\cancel{1,-1},\tilde{j}_{k-3})\\
	&=(1,1,\tilde{i}_{k-3})\star(-1,1,\tilde{j}_{k-3})7=(\G_{\frac{d}{4},2^{k-3}})_{i,j}
\end{align*}
Hence $N_{1,1}=N_{2,2}$ and $N_{1,2}=N_{2,1}=\G_{\frac{d}{4},2^{k-3}}$.
Therefore we have $A^{(d)}_{2^{k-2}}=\begin{pmatrix}A^{(d)}_{2^{k-3}}&\G_{\frac{d}{4},2^{k-3}}\\\G_{\frac{d}{4},2^{k-3}}&A^{(d)}_{2^{k-3}}\end{pmatrix}$ for some $(2^{k-3}\times 2^{k-3})$ matrix $A_{2^{k-3}}^{(d)}$. Next, we continue to partition $A^{(d)}_{2^{k-3}}$ into four sub square blocks as $A^{(d)}_{2^{k-3}}=\begin{pmatrix}
	N'_{1,1}&N'_{1,2}\\N'_{2,1}&N'_{2,2}
\end{pmatrix}$. For $i,j\in\{1,...,2^{k-4}\}$, we have
\begin{align*}
	(N'_{1,2})_{i,j}&=(1,1,1,1,1,\tilde{i}_{k-4})\star(-1,1,1,1,-1,\tilde{j}_{k-4})\\
 &=(1,1,1,1,-1,\tilde{i}_{k-4})\star(-1,1,1,1,1,\tilde{j}_{k-4})=(N'_{2,1})_{i,j}\\
	(N'_{1,1})_{i,j}&=(1,1,1,1,1,\tilde{i}_{k-4})\star(-1,1,1,1,1,\tilde{j}_{k-4})\\
 &=(1,1,1,1,-1,\tilde{i}_{k-4})\star(-1,1,1,1,-1,\tilde{j}_{k-4})=(N'_{2,2})_{i,j}\\
	(N'_{1,2})_{i,j}&=(1,1,1,1,1,\tilde{i}_{k-4})\star(-1,1,1,1,-1,\tilde{j}_{k-4})7=(1,1,1,\cancel{1,1},\tilde{i}_{k-4})\star(-1,1,1,\cancel{1,-1},\tilde{j}_{k-4})\\
	&=(1,1,1,\tilde{i}_{k-4})\star(-1,1,1,\tilde{j}_{k-4})7=(\G_{\frac{d}{4},2^{k-4}})_{i,j}
\end{align*}
Hence $N'_{1,1}=N'_{2,2}$ and $N'_{1,2}=N'_{2,1}=\G_{\frac{d}{4},2^{k-4}}$.
Therefore we have $A^{(d)}_{2^{k-3}}=\begin{pmatrix}A^{(d)}_{2^{k-4}}&\G_{\frac{d}{4},2^{k-4}}\\\G_{\frac{d}{4},2^{k-4}}&A^{(d)}_{2^{k-4}}\end{pmatrix}$ for some $(2^{k-4}\times 2^{k-4})$ matrix $A_{2^{k-4}}^{(d)}$. We will continue the partition procedure (i.e., partition $A_{2^{k-i}}^{(d)}$ for $i\in\{2,...,k-1\}$) and apply the same argument to obtain
\[A^{(d)}_{2^{k-i}}=\begin{pmatrix}A^{(d)}_{2^{k-i-1}}&\G_{\frac{d}{4},2^{k-i-1}}\\\G_{\frac{d}{4},2^{k-i-1}}&A^{(d)}_{2^{k-i-1}}\end{pmatrix}\]
where $i\in\{2,...k-1\}$. For $A_1^{(d)}$, its entry equals to $(\G_{2^k})_{1,1}=(1,...,1)\star(-1,1,...,1)=1$.

Next, we consider $B^{(d)}_{2^{k-2}}$. First, we partition $B^{(d)}_{2^{k-2}}$ into four sub blocks as $B^{(d)}_{2^{k-2}}=\begin{pmatrix}
	N_{1,1}&N_{1,2}\\N_{2,1}&N_{2,2}
\end{pmatrix}$. For $i,j\in\{1,...,2^{k-3}\}$, we have
\begin{align*}
	(N_{1,2})_{i,j}&=(1,1,1,1,\tilde{i}_{k-3})\star(-1,-1,-1,-1,\tilde{j}_{k-3})\\
 &=(1,1,1,-1,\tilde{i}_{k-3})\star(-1,-1,-1,1,\tilde{j}_{k-3})=(N_{2,1})_{i,j}\\
	(N_{1,1})_{i,j}&=(1,1,1,1,\tilde{i}_{k-3})\star(-1,-1,-1,1,\tilde{j}_{k-3})\\
 &=(1,1,1,-1,\tilde{i}_{k-3})\star(-1,-1,-1,-1,\tilde{j}_{k-3})=(N_{2,2})_{i,j}\\
	(N_{1,1})_{i,j}&=(1,1,1,1,\tilde{i}_{k-3})\star(-1,-1,-1,1,\tilde{j}_{k-3})\\
	&=(1,1,\cancel{1,1},\tilde{i}_{k-3})\star(-1,-1,\cancel{-1,1},\tilde{j}_{k-3})=(\overline{\G_{\frac{d}{4},2^{k-3}}})_{i,j}
\end{align*}
Thus we have $N_{1,1}=N_{2,2}=\overline{\G_{\frac{d}{4},2^{k-3}}}$ and $N_{1,2}=N_{2,1}$. Hence we can conclude that $B^{(d)}_{2^{k-2}}=\begin{pmatrix}\overline{\G_{\frac{d}{4},2^{k-3}}}&B^{(d)}_{2^{k-3}}\\B^{(d)}_{2^{k-3}}&\overline{\G_{\frac{d}{4},2^{k-3}}}\end{pmatrix}$ for some $(2^{k-3}\times 2^{k-3})$ matrix $B^{(d)}_{2^{k-3}}$. Next, we continue to partition $B^{(d)}_{2^{k-3}}$ into four blocks as $B^{(d)}_{2^{k-3}}=\begin{pmatrix}N'_{1,1}&N'_{1,2}\\ N'_{2,1}&N'_{2,2}\end{pmatrix}$. For $i,j\in\{1,...,2^{k-4}\}$, we have
\begin{align*}
	(N'_{1,2})_{i,j}&=(1,1,1,1,1,\tilde{i}_{k-4})\star(-1,-1,-1,-1,-1,\tilde{j}_{k-4})\\
	&=(1,1,1,1,-1,\tilde{i}_{k-4})\star(-1,-1,-1,-1,1,\tilde{j}_{k-4})=(N'_{2,1})_{i,j}\\
	(N'_{1,1})_{i,j}&=(1,1,1,1,1,\tilde{i}_{k-4})\star(-1,-1,-1,-1,1,\tilde{j}_{k-4})\\
	&=(1,1,1,1,-1,\tilde{i}_{k-4})\star(-1,-1,-1,-1,-1,\tilde{j}_{k-4})=(N'_{2,2})_{i,j}\\
	(N'_{1,1})_{i,j}&=(1,1,1,1,1,\tilde{i}_{k-4})\star(-1,-1,-1,-1,1,\tilde{j}_{k-4})\\
	&=(1,1,1,\cancel{1,1},\tilde{i}_{k-4})\star(-1,-1,-1,\cancel{-1,1},\tilde{j}_{k-4})=(\overline{\G_{\frac{d}{4},2^{k-4}}})_{i,j}
\end{align*}
Thus we have $N'_{1,1}=N'_{2,2}=\overline{\G_{\frac{d}{4},2^{k-4}}}$ and $N'_{1,2}=N'_{2,1}$. Hence we can conclude that $B^{(d)}_{2^{k-3}}=\begin{pmatrix}\overline{\G_{\frac{d}{4},2^{k-4}}}&B^{(d)}_{2^{k-4}}\\B^{(d)}_{2^{k-4}}&\overline{\G_{\frac{d}{4},2^{k-4}}}\end{pmatrix}$ for some $(2^{k-4}\times 2^{k-4})$ matrix $B^{(d)}_{2^{k-4}}$. We will continue the partition procedure (i.e., partition $B_{2^{k-i}}^{(d)}$ for $i\in\{2,...,k-1\}$) and apply the same argument to obtain
$$B^{(d)}_{2^{k-i}}=\begin{pmatrix}\overline{\G_{\frac{d}{4},2^{k-i-1}}}&B^{(d)}_{2^{k-i-1}}\\B^{(d)}_{2^{k-i-1}}&\overline{\G_{\frac{d}{4},2^{k-i-1}}}\end{pmatrix}$$
where $i\in\{2,...,k-1\}$. For $B_1^{(d)}$, its entry equals to $(\G_{2^k})_{1,2^k}=(1,...,1)\star(-1,...,-1)=-1$.
\end{proof}

\begin{definition}
The Hadamard matrix $H_k$ is a $2^k$ by $2^k$ square matrix defines as follows:
\[H_1=\frac{1}{\sqrt{2}}\begin{pmatrix}1&1\\1&-1\end{pmatrix}\]
and
\[H_k=\frac{1}{\sqrt{2}}\begin{pmatrix}H_{k-1}&H_{k-1}\\H_{k-1}&-H_{k-1}\end{pmatrix}\]
\end{definition}
\begin{lemma}{\label{Alem_diag}}
Let $M=\begin{pmatrix}A&B\\B&A\end{pmatrix}$ where $A$ and $B$ be a $2^k$ by $2^k$ square matrix. If $A$ and $B$ are diagonal in the Hadamard basis, then $M$ is diagonal in the Hadamard basis.
\end{lemma}
\begin{proof}
By direct calculation, we have 
$$H_{k+1}\begin{pmatrix}A&B\\B&A\end{pmatrix}H_{k+1}^{-1}=\begin{pmatrix}H_k(A+B)H_k^{-1}&0\\0&H_k(A-B)H_k^{-1}\end{pmatrix}$$
Thus if $A$ and $B$ are diagonal in the Hadamard basis, then $M$ is also diagonal in the Hadamard basis.
\end{proof}

\begin{lemma}{\label{Alem_diag_game}}
Consider the game matrix $\G_{2^k}$ where $k\geq 4$.

(1) If $\G_{2^{k-2},2^{i}}$ are diagonal in the Hadamard basis for all $i\in\{0,...,k-2\}$, then $A_{2^j}^{(2^k)}$ are diagonal in the Hadamard basis for all $j\in\{0,...,k-2\}$. Furthermore, we have $\G_{2^k,2^{j}}$ are diagonal in the Hadamard basis for all $j\in\{0,...,k-1\}$

(2) If $\overline{\G_{2^{k-2},2^{i}}}$ are diagonal in the Hadamard basis for all $i\in\{0,...,k-2\}$, then $B_{2^j}^{(2^k)}$ are diagonal in the Hadamard basis for all $j\in\{0,...,k-2\}$. Furthermore, we have $\overline{\G_{2^k,2^{j}}}$ are diagonal in the Hadamard basis for all $j\in\{0,...,k-1\}$ 
\end{lemma}
\begin{proof}
Let $d=dim(\G_{2^k})=2^k$. Recall that
$$A^{(d)}_{2^{k-2}}=\begin{pmatrix}A^{(d)}_{2^{k-3}}&\G_{\frac{d}{4},2^{k-3}}\\\G_{\frac{d}{4},2^{k-3}}&A^{(d)}_{2^{k-3}}\end{pmatrix}$$
Observe that $\G_{2^k,2^j}=A_{2^j}^{(d)}$ for all $j\in\{0,...,k-2\}$.
By Lemma \ref{Alem_diag}, $A_{2}^{(2^k)}=\begin{pmatrix}1&\G_{2^{k-2},1}\\ \G_{2^{k-2},1}&1\end{pmatrix}$
is diagonal in the Hadamard basis. By Lemma \ref{Alem_diag} again, since $A_{2}^{(2^k)}$ and $\G_{2^{k-2},2}$ are diagonal in the Hadamard basis, matrix $A^{(2^k)}_{2^2}=\begin{pmatrix}A^{(2^k)}_{2}&\G_{2^{k-2},2}\\\G_{2^{k-2},2}&A^{(2^k)}_{2}\end{pmatrix}$ is also diagonal in the Hadamard basis. Inductively, we can conclude that $A_{2^j}^{(2^k)}=\G_{2^k,2^j}$ are diagonal in the Hadamard basis for all $j\in\{0,...,k-2\}$. Since $A_{2^{k-2}}^{(2^k)}$ and $\G_{2^{k-2}}$ are diagonal in the Hadamard basis, by Lemma \ref{Alem_diag}, matrix $\G_{2^k,2^{k-1}}=\begin{pmatrix}A_{2^{k-2}}^{(2^k)}&\G_{2^{k-2}}\\ \G_{2^{k-2}}& A_{2^{k-2}}^{(2^k)}\end{pmatrix}$ is also diagonal in the Hadamard basis. Thus statement (1) is proved.

Recall that 
$$B^{(d)}_{2^{k-2}}=\begin{pmatrix}\overline{\G_{\frac{d}{4},2^{k-3}}}&B^{(d)}_{2^{k-3}}\\B^{(d)}_{2^{k-3}}&\overline{\G_{\frac{d}{4},2^{k-3}}}\end{pmatrix}$$
Observe that $\overline{\G_{2^k,2^j}}=B_{2^k}^{(d)}$ for all $j\in\{0,...,k-2\}$. By Lemma \ref{Alem_diag}, $B_2^{(2^k)}=\begin{pmatrix}\overline{\G_{2^{k-2},1}}&-1\\-1&\overline{\G_{2^{k-2},1}}\end{pmatrix}$ is diagonal in the Hadamard basis. By Lemma \ref{Alem_diag} again, since $\overline{\G_{2^{k-2},2}}$ and $B^{(2^{k})}_{2}$ are diagonal in the Hadamard basis, matrix $B^{(2^k)}_{2^2}=\begin{pmatrix}\overline{\G_{2^{k-2},2}}&B^{(2^{k})}_{2}\\B^{(2^{k})}_{2}&\overline{\G_{2^{k-2},2}}\end{pmatrix}$ is also diagonal in the Hadamard basis. Inductively, we can conclude that $\overline{\G_{2^k,2^j}}=B_{2^k}^{(d)}$ are diagonal in the Hadamard basis for all $j\in\{0,...,k-2\}$. Since $\G_{2^{k-2}}$ and $B_{2^{k-2}}^{(d)}$ are diagonal in the Hadamard basis, by Lemma \ref{Alem_diag}, matrix $\overline{\G_{2^k,2^{k-1}}}=\begin{pmatrix}\G_{2^{k-2}}&B_{2^{k-2}}^{(d)}\\B_{2^{k-2}}^{(d)}& \G_{2^{k-2}}\end{pmatrix}$ is also diagonal in the Hadamard basis. Thus statement (2) is proved.
\end{proof}

\begin{corollary}
Let $k\geq 2$.
Then $\G_{2^k,2^i}$ and $\overline{\G_{2^k,2^i}}$ are diagonal in the Hadamard basis for all $i\in\{0,...,k\}$. In particular, the game matrix $\G_{2^k}$ is diagonal in the Hadamard basis.
\end{corollary}
\begin{proof}
We will proceed by induction. For the induction step. Assume $\G_{2^k,2^i}$ and $\overline{\G_{2^k,2^i}}$ are diagonal in the Hadamard basis for all $i\in\{0,...,k\}$. By Lemma \ref{Alem_diag_game}, $\G_{2^{k+2},2^i}$ and $\overline{\G_{2^{k+2},2^i}}$ are diagonal in the Hadamard basis for all $i\in\{0,...,k+1\}$. Observe that we can express $\G_{2^{k+2}}$ as
$$\G_{2^{k+2}}=\begin{pmatrix}\G_{2^{k+2},2^{k+1}}&\overline{\G_{2^{k+2},2^{k+1}}}\\\overline{\G_{2^{k+2},2^{k+1}}}&\G_{2^{k+2},2^{k+1}}\end{pmatrix}$$
By Lemma \ref{Alem_diag_game}, since $\G_{2^{k+2},2^{k+1}}$ and $\overline{\G_{2^{k+2},2^{k+1}}}$ are diagonal in the Hadamard basis, we can conclude that $\G_{2^{k+2}}$ is also diagonal in the Hadamard basis, which finishes the induction step. The base cases are $k=2$ and $k=3$. Recall the matrix $\G_4$ and $\G_8$ are shown in Example \ref{Aex}. We can check that for $k\in\{2,3\}$, matrices
$\G_{2^{k},2^i}$ and $\overline{\G_{2^{k},2^i}}$ are diagonal in the Hadamard basis for all $i\in\{0,...,k\}$.
\end{proof}

\end{document}